\documentclass[prb,showpacs,preprintnumbers,amsmath,aps]{revtex4}
\usepackage{graphicx,hyperref}
\usepackage{amssymb}
\usepackage{bm}
\usepackage{subfigure}
\usepackage{color}

\let\a=\alpha \let\b=\beta  \let\g=\gamma  \let\d=\delta \let\e=\varepsilon
       
\let\m=\mu                 \let\r=\rho
\let\s=\sigma    \let\f=\varphi  

\def\ie{{\it i.e., }}
\def\eg{{\it e.g., }}

\newcommand{\beq}{\begin{equation}}
\newcommand{\eeq}{\end{equation}}
\newcommand{\bea}{\begin{eqnarray}}
\newcommand{\eea}{\end{eqnarray}}
\newcommand{\beal}{\begin{aligned}}
\newcommand{\eal}{\end{aligned}}

\newcommand{\dCaCbi}{\delta_{\mathcal{C}_i^a,\mathcal{C}_i^b}}

\newcommand{\sumiz}{\sum_{i}^{(0)}}

\newcommand{\sumiuz}{\sum_{i}^{(1)+(0)}}
\newcommand{\sumidz}{\sum_{i}^{(2)+(0)}}
\newcommand{\sumabu}{\sum_{a \neq b}^{(1)}}
\newcommand{\sumabd}{\sum_{a \neq b}^{(2)}}
\newcommand{\sumau}{\sum_{a}^{(1)}}
\newcommand{\sumad}{\sum_{a}^{(2)}}
\newcommand{\sumbd}{\sum_{b}^{(2)}}

\begin{document}

\title{Random-Field Ising like effective theory of the glass transition: I Mean-Field Models}

\author{Giulio Biroli} \email{giulio.biroli@cea.fr}
\affiliation{IPhT, CEA/DSM-CNRS/URA 2306, CEA Saclay, F-91191 Gif-sur-Yvette Cedex, France\\
Laboratoire de Physique Statistique, \'Ecole Normale Sup\'erieure, CNRS, France}

\author{Chiara Cammarota} \email{chiara.cammarota@kcl.ac.uk}
\affiliation{Department of Mathematics, King's College London, Strand, London WC2R 2LS, UK}

\author{Gilles Tarjus} \email{tarjus@lptmc.jussieu.fr}
\affiliation{LPTMC, CNRS-UMR 7600, Sorbonne Universit\'e, 4 Pl. Jussieu, F-75005 Paris, France}

\author{Marco Tarzia} \email{tarzia@lptmc.jussieu.fr}
\affiliation{LPTMC, CNRS-UMR 7600, Sorbonne Universit\'e, 4 Pl. Jussieu, F-75005 Paris, France}

\date{\today}

\begin{abstract}
In this paper and in the companion one\cite{paper2} we address the problem of identifying the effective theory that describes the statistics of the fluctuations of what is thought to be the relevant order parameter for glassy systems---the overlap field with an equilibrium reference configuration---close to the putative thermodynamic glass transition. Our starting point is the mean-field theory of glass formation which relies on the existence of a complex free-energy landscape with a multitude of metastable states. In this paper, we focus on archetypal mean-field models possessing this type of free-energy landscape and set up the framework to determine the exact effective theory. We show that the effective theory at the mean-field level is generically of the random-field + random-bond Ising type.
We also discuss what are the main issues concerning the extension of our result to finite-dimensional systems. This extension is addressed in detail in the companion paper.   
\end{abstract}

\maketitle

\tableofcontents

\section{Introduction} 
\label{sec:intro}

Developing a proper treatment of nonperturbative fluctuations is one of the most difficult methodological issues one can encounter when facing a physical problem.  In the past, this difficulty has been circumvented mainly in two ways: either by making use of very clever assumptions on the physics of the problem at hand  or by finding a mapping that transforms the original strong-coupling model with nonperturbative fluctuations (that one cannot solve) in a weak-coupling model with perturbative fluctuations (that one can treat easily). Examples of the first case are provided by several variational wave-functions discovered along the years in condensed-matter physics ({\it e.g.}, the Laughlin wave-function for the fractional quantum Hall effect\cite{laughlin}).  Examples of the latter are provided by the study of the low-temperature behavior of systems that are characterized by dilute nonperturbative excitations, such as the XY model that can be mapped onto a dilute Coulomb gas of vortices.\cite{vortices}  Actually, in all of these situations one solves the problem by {\it avoiding} to directly tackle nonperturbative fluctuations and by instead finding a suitable short-cut. There are however cases, as for example the glass transition of supercooled liquids,\cite{Wolynesbook,GiulioLudo,Pedestrians,MCT_review,Glass_review,Tarjus_review} where these two approaches seem to fail: No dual weak-coupling system can be identified, no clear-cut assumptions to simplify the problem can be made.  Two theoretical approaches of glass formation, the dynamical-facilitation theory~\cite{Chandler} and the approach based on geometrical frustration and avoided criticality,\cite{Tarjus1} provide valuable attempts to identify and address the source of relevant nonperturbative fluctuations in glass-forming liquids,  but they remain at present not fully satisfactory.

Given this situation, an alternative route is to start from an established mean-field description and to incorporate, up to some finite length scale, the fluctuations of the identified order parameter in an {\it effective theory}. The hope is to derive an effective theory that (i) encompasses the main physical ingredients while leaving out inessential ones and (ii) contains nonperturbative fluctuations that can be handled in a more tractable manner than in the original problem. The aim of this work is to perform the first steps toward such an effective theory of the glass transition. 

Our starting point is the mean-field theory of glasses\cite{KTW,RFOT_review} which has recently gained momentum through the solution of the hard-sphere glass in infinite dimensions.\cite{kurchan-zanponi-etal,kurchan-maimbourg} The associated scenario relies on the existence, below a critical temperature associated with a dynamical transition, of a complex free-energy landscape with a multitude of metastable states that is characterized by an extensive configurational entropy. An ideal thermodynamic glass transition, known as a random first-order transition (RFOT), takes place when the configurational entropy becomes subextensive.\cite{KTW} The relevant order parameter is then provided by the similarity or overlap between equilibrium liquid configurations.\cite{franzparisi-potential,franzparisi-potential2,franzparisi-potential3} However, this mean-field scenario appears fragile to the introduction of fluctuations,\cite{BB-review,BCTT-PRB} and the very notion of metastable states is well-defined only when fluctuations are absent, as in a mean-field approximation, or suppressed, as in a small system.\cite{rulquin}

Developing an effective theory of glass-forming systems directly formulated in terms of what is thought to be the physically relevant local order parameter, the overlap with an equilibrium reference configuration, seems a valuable task for several conceptual and technical reasons:

(a) It provides a more intuitive description  of the glass transition and, most importantly, allows one to circumvent the explicit description in terms of metastable states.

(b) The problem of handling in a fully satisfactory way the large-scale physics described by the replica field theory suggested by mean-field models
remains very challenging, despite some recent theoretical progress ({\it e.g.}, instantons calculations\cite{Dzero}, Kac analysis\cite{Franz_Kac}, and real space RG approaches\cite{tarzia-cammarota-etal,Castellana,Angelini}). This is partly due to the complicated replica matrix structure of the overlap fields. Focusing only on some of the overlaps, namely those involving the equilibrium reference configuration, while integrating out all the others naturally leads to a scalar field theory in the presence of quenched disorder. The latter is {\it a priori} much easier to handle than the original theory and can be studied by using powerful tools of statistical physics ({\it e.g.}, large-scale numerical simulations, nonperturbative functional renormalization group, etc.).

(c) Promoting the order parameter to a fully fluctuating field is a way to study fluctuations and correlations beyond mean-field theory. This provides a proper description of all large-scale and nonperturbative fluctuations and thereby allows one to assess the nature of the critical points and  identify the mechanisms that  could possibly destroy or alter the glass transition (RFOT) in finite dimensions.\cite{Moore}

This program has already been partly achieved. A mapping to an effective theory akin to the Ising model in a random field (RFIM)\cite{Nattermann} has been derived near (but below) the dynamical transition of the mean-field theory\cite{silvio_mct} as well as near the critical points appearing in an extended phase diagram in the presence of additional sources or pinning fields.\cite{silvio_Tc,Franz_pinning,noi_Tc} In this work we focus on the more challenging problem of establishing an effective theory in the vicinity of the putative thermodynamic glass transition.

In this first paper, we consider two archetypal mean-field models for glass formation, {\it i.e.}, the  Random Energy Model\cite{derrida} (REM) and its Kac-like generalization to a finite number ($2^M$) of states per site\cite{KREM} (which we call in the following the $2^M$-KREM) on a fully connected lattice. In the REM case
we show that the statistics of the thermal fluctuations of the global overlap with an equilibrium configuration is {\it exactly} described by an Ising variable $\sigma = \pm 1$ subjected to a random field, {\it i.e} a $0$-dimensional RFIM,
\begin{equation} \label{eq:Heff_REM}
\beta {\cal H}_{\rm eff} = {\cal S}_0-(H + \delta h) \sigma \, ,
\end{equation}
where the two values of $\sigma$ correspond to one or zero overlap, $H$ is a temperature-dependent uniform field of order $N$ that corresponds to the configurational entropy and vanishes at the RFOT temperature $T_K$, and $\delta h$ a random field of zero mean and fluctuations of order $\sqrt{N}$, where $N \to \infty$ is the logarithm of the number of states. A similar (but richer) result can be obtained for the fully connected $2^M$-KREM, for which one can show that the effective theory for the 
overlap profile near $T_K$ corresponds to a fully connected random-bond+random-field Ising model with multi-body interactions and higher order random terms: 
\begin{equation} \label{eq:Heff_REMFC}
	\beta {\cal H}_{\rm eff} = {\cal S}_0 -\sum_i (H + \delta h_i) \sigma^i - \frac 1 2 \sum_{i \neq j} \Big( \frac{J_2}{N} + \frac{\delta J_{2,ij}}{\sqrt{N}} \Big) \sigma^i \sigma^j
-  \frac{J_3}{3!N^2} \sum_{i , j, k \neq} \sigma^i \sigma^j \sigma^k  - \frac{J_4}{4!N^3} \sum_{i , j, k,l \neq} \sigma^i \sigma^j \sigma^k \sigma^l + \cdots \, .
\end{equation}
The $2$ possible states $\sigma^i=\pm 1$ correspond to a low- and a high-overlap with a reference equilibrium configuration on a given site $i$; 
the field $H$ plays the role of the  configurational entropy, vanishing at $T_K$, $\delta J_{ij}$ and $\delta h_i$ are random variables with zero mean;  $J_2>0$ is a ferromagnetic coupling and $J_3$ and $J_4$ are $3$- and $4$-body interactions (whose sign can depend on the parameters of the original microscopic model, \eg the number of states).  The ellipses denote multi-body  interactions beyond the $4$-body one and higher-order random terms which have been omitted. The  coupling constants and the variance of the random terms can be, at least in principle, computed exactly.

The REM and the fully connected $2^M$-KREM can of course be exactly solved, with no need to go through a mapping onto an effective Hamiltonian, but the present treatment illustrates how an effective Ising theory with quenched disorder emerges and this sets the stage for studying finite-dimensional glass-formers. The latter, including glass-forming liquids, will be the focus of  the companion paper\cite{paper2}. In this case additional approximations are required but the output will again be a description of the glass transition in terms of an Ising model in an external field with random-field and random-bond disorder and  long-range competing multi-body interactions.

The rest of the paper is organized as follows. In Sec.~\ref{sec:known} we review the situations where the random-field Ising model in one form or another appears in the theory of glass-forming liquids and we provide a general intuitive argument for why this is so. The following section, Sec.~\ref{sec:mean_field}, is devoted to the derivation of the effective theory for the overlap with a reference equilibrium configuration in two mean-field models of structural glasses, the REM and its generalization to a finite number of states, the  $2^M$-KREM. We focus on the region around the putative thermodynamic glass transition (RFOT). In Sec.~\ref{discussion} we first illustrate the difficulties that one encounters when trying to generalize the procedure developed for the mean-field models to finite-dimensional glass-forming systems. (This generalization will be the topic of the companion paper.\cite{paper2}) We next discuss some of the properties of the quenched disorder appearing in the effective theory. Finally, some concluding remarks are given in Sec.~\ref{sec:conclusions}. Most of the technical details of the calculations are presented in two Appendices.

\section{RFIM-like criticality in glass-forming liquids}
\label{sec:known}

\subsection{Known results} 

\begin{figure}
\centering
\includegraphics[width=0.64\textwidth]{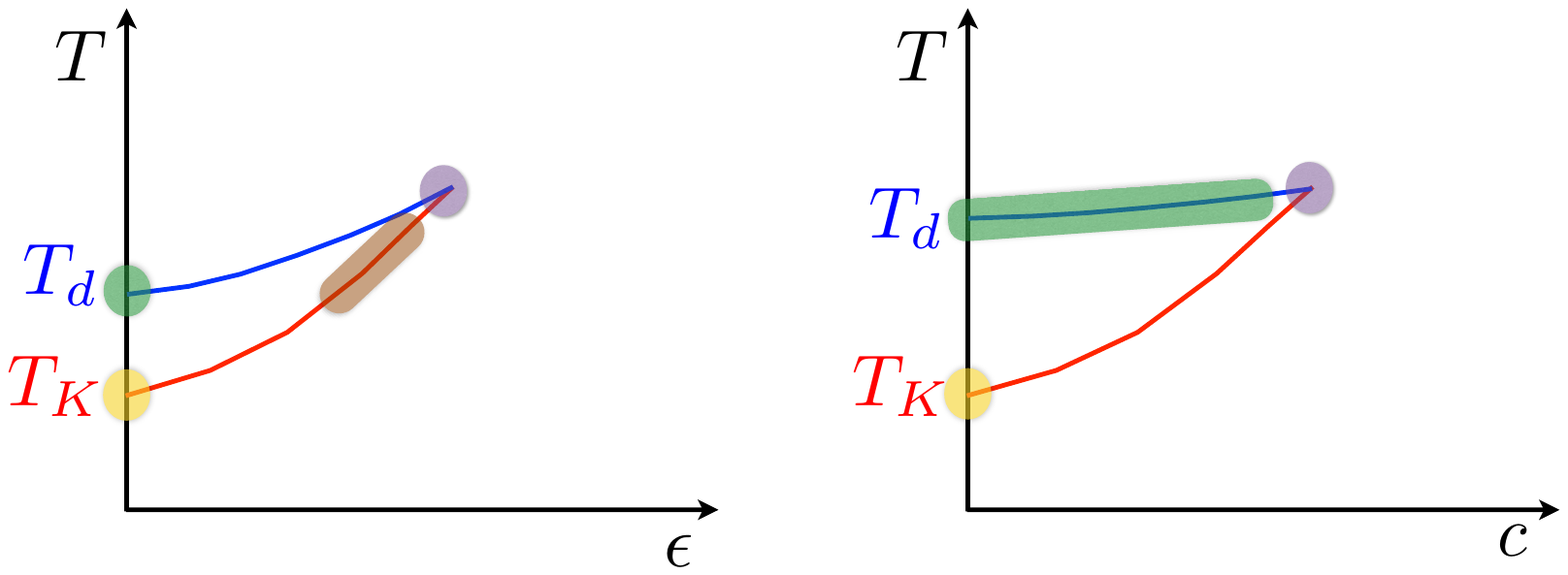}
\vspace{-3.8cm}
\caption{Sketch of the phase diagram of constrained glass-forming systems predicted by the mean-field/RFOT theory in the ($T$-$\epsilon$) plane (left) and in the ($T$-$c$) plane (right), showing the regions where RFIM-like universality classes of the critical points in finite dimensions (see text for a more detailed description). The dynamical (MCT) glass transitions are in the same universality class as the spinodal point of the RFIM\cite{silvio_mct} (green). The terminal critical points in the ($T$-$\epsilon$) plane and in the ($T$-$c$) plane belong to the universality class of the RFIM\cite{silvio_Tc,noi_Tc,Cammarota_pinning} (violet). The first-order transition line in the ($T$-$\epsilon$) plane is described by a first-order transition in the presence of a random field\cite{noi_Tc} (brown). The situation in the absence of coupling ($\epsilon=0$, $c=0$) and close to the putative RFOT $T_K$ (yellow) has been investigated for the first time numerically in Ref.~[\onlinecite{Stevenson}] and is the focus of the present paper. Note that one expects that the putative line of RFOT in the ($T$-$c$) plane is described by a similar effective theory as that near $T_K$.}
\label{fig:RFIM}
\end{figure}

This is not the first time that an Ising model in a random field appears in the context of supercooled liquids. Actually, the idea of mapping the glass transition onto a magnetic system with quenched disorder was put forward and analyzed for the first time in Ref.~[\onlinecite{Stevenson}]. Several recent analytical and numerical investigations strongly support the idea that the effective theory which describes the thermal fluctuations of the overlap with an equilibrium configuration in glassy systems is provided by an Ising model in the presence of quenched disorder. Below we present a list of the main known results, which are pictorially summarized in Fig.~\ref{fig:RFIM}.

From the analysis of the perturbation theory of the replica field description it was first shown in Ref.~[\onlinecite{silvio_mct}] that the critical fluctuations of the overlap close to the dynamical (mode-coupling-like) transition, in the so-called $\beta$-regime just below the transition, are in the same universality class as those found at the spinodal point of the (standard, short-range) RFIM. Both types of singularities, the dynamical transition and the spinodal, can only be present when activated events such as nucleation are not taken into account. This connection was further examined in Ref.~[\onlinecite{Nandi}], where the spinodal of the RFIM  was studied  at zero temperature, thereby eliminating all thermal fluctuations. 

In the past few years some effort has been devoted to analyze the universality class of the critical points that can be induced in glassy systems by the
presence of suitable constraints. The first such case that was studied corresponds to introducing an additional attractive coupling $\epsilon$ to a reference equilibrium configuration of the system which in effect acts as a source linearly coupled to the overlap with this reference configuration. Within the mean-field theory, the glass transition found at $T = T_K$ and $\e = 0$ transforms into a line of first-order transition in the ($T$-$\epsilon$) plane, which ends in a critical point at $T=T_c$ and $\e = \e_c >0$, as illustrated in left panel of Fig.~\ref{fig:RFIM}. This feature is a key prediction of the mean-field/RFOT theory. It is found in mean-field disordered spin models,\cite{franzparisi-potential3} and evidence for it has been observed in computer simulations of $3$-dimensional atomistic models.\cite{franzparisi-potential2,franzparisi-potential3,Kob,Berthier2,Parisi_Bea,Berthier3,Berthier-swap} Through a thorough analysis of the soft modes emerging at the terminal critical point in the replica field theory and of the resulting properties of the perturbation theory,\cite{silvio_Tc} it was established that this critical point belongs to the universality class of the standard RFIM in finite dimensions. We also found the same result independently, by using an  approach based on an expansion in increasing number of free replica sums.\cite{noi_Tc} Of course this is valid if the transition is not destroyed by the disorder, but numerical indications that the RFIM critical behavior can indeed be found in a $3$-dimensional glass-forming liquid model has been recently obtained.\cite{Berthier3} (Finally, a further link between the physics of RFIM and supercooled liquids comes from 
the study of fluctuations of amorphous interfaces in $3$-dimensional liquid models\cite{roughinterfaces} whose statistical properties have also been investigated through an expansion in free replica sums.\cite{GCPRX})

Surprisingly enough, the relevance of this RFIM-like criticality to glass-forming systems also appears in the context of plaquette spin models, usually taken as an illustration of the dynamical-facilitation theory of glass formation.\cite{ritort-sollich,garrahan02} In a series of papers\cite{Garrahan2,Jack1,Turner} Garrahan and coworkers analyzed plaquette spin models in dimensions $d= 2$ (the ``triangular plaquette model'') and $d = 3$ (the ``square pyramid model''), focusing on the thermodynamic behavior in the presence of an attractive coupling $\e$. For the $3d$ square pyramid model the authors presented strong numerical evidence in favor of the existence of a transition line in the ($T$-$\e$) plane, terminating in a critical point whose universal properties are those of the $3d$ RFIM. On the other hand, no such transition was found in $d=2$, in agreement with the fact that the lower critical dimension of the RFIM is precisely $d=2$. The role of short- versus long-range fluctuations of the overlap was also studied in these models by means of Bethe-lattice calculations.\cite{rulquin-plaquettes}

Another procedure to constrain the system toward a reference configuration (referred to as the ``pinned particles'' method) is to freeze the positions of a randomly chosen fraction $c$ of the particles to the values they have in a given equilibrium configuration.\cite{Cammarota_pinning} According to the mean-field/RFOT theory, the constraint induces a line $c_K(T)$ of thermodynamic glass transition (RFOT) and a line of dynamical (mode-coupling-like) glass transition $c_d(T)$ in the ($T$-$c$) plane, as illustrated in the right panel of Fig.~\ref{fig:RFIM}. Differently from the case of the $\e$-coupling where a nonzero $\epsilon$ transforms the thermodynamic (RFOT)  glass transition into a conventional first-order transition (albeit in the presence of a random field\cite{noi_Tc}), in the pinned-particle case the thermodynamic glass transition keeps its glassy RFOT character all along the line. The RFOT line and the line of dynamical transitions merge in a critical endpoint. This scenario is realized in mean-field models of glass-forming systems,\cite{Cammarota_pinning,Ricci,Cammarota1} and its relevance for finite dimensions is supported by calculations based on a Migdal-Kadanoff real-space renormalization group (RG)\cite{Cammarota_pinning} and numerical results.\cite{Kobp,Kob,BerthierKob} In Ref.~[\onlinecite{Franz_pinning}] it was established that, just like for the case of the $\epsilon$-coupling, the critical endpoint in the ($T$-$c$) plane is in the same universality class as the critical point of the RFIM, also in agreement with the real-space RG results of Ref.~[\onlinecite{Cammarota_pinning}]. Moreover, the mode-coupling theory (MCT) predicts several kinds of critical dynamical behavior for randomly pinned systems.\cite{Cammarota2,Krakoviack,Szamel} Along the dynamical transition line, $c_d(T)$, the transition remains of $A_2$ type in the MCT terminology,\cite{MCT_review}) until the terminal point is reached, where the singularity becomes of $A_3$ type. The dynamical behavior at the $A_3$ critical endpoint displays activated dynamical scaling, a characteristic property of the critical dynamical behavior of the RFIM.\cite{Nandi1}

Finally, the dynamics of the kinetically constrained Fredrickson-Andersen model\cite{Frederickson} was analyzed on a Bethe lattice, showing the presence of a dynamical transition whose finite size-scaling is consistent with that of the RFIM.\cite{Franz_Sellitto} The Fredrickson-Andersen model on a Bethe lattice was also studied in the presence of a random pinning.\cite{Ikeda} Strong evidence was then found for the existence of a line $c_d(T)$ of dynamical glass transitions with the characteristic $A_2$ MCT singularity ending in a critical point with an $A_3$ MCT singularity related to the dynamical behavior of the RFIM (see above).


The mapping of the properties of glass-forming systems to those of the RFIM obtained so far has two main limitations. The first one is that it concerns the fluctuations of the overlap field but not directly the dynamical behavior, {\it i.e.} there is no direct connection between the {\it dynamics} of the RFIM and that of supercooled liquids. Recently, Rizzo \cite{rizzo} went beyond the static analysis to obtain a dynamical stochastic equation, called Stochastic-Beta-Relaxation equation, involving the effect a random field. This provides a theory of dynamical fluctuations in finite dimensions close to the avoided dynamical transition (MCT crossover) but is far from providing a full dynamical description of the approach to the physical glass transition. The second limitation is that the mapping, summarized in Fig.~\ref{fig:RFIM},  has so far left aside the more interesting, and more challenging, case that corresponds to the situation in the absence of coupling ($\e=0$, $c=0$) and close to $T_K$, where a thermodynamic glass transition of RFOT type is predicted at the mean-field level. Whereas we do not address the first point in this work, we  do consider the second issue of the mapping in the vicinity of the putative thermodynamic glass transition.

\subsection{Self-induced disorder in the overlap field theory}
\label{sec:self-induced}

An intuitive argument explaining why the effective theory describing the local fluctuations of the overlap order parameter in glass-forming systems is in the class of an Ising model in a random field relies on the concept of ``self-induced disorder''.\cite{silvio_mct} In fact, the object which plays the role of a (large deviation) Landau-like functional for the chosen order parameter is the average free-energy cost that is necessary to keep the system at an overlap $p({\bf x})$ with a  reference equilibrium configuration $\mathcal{C}_{\rm eq}$. To be more concrete, take a glass-former described by configurations ${\cal C}$ and a Hamiltonian ${\cal H}[{\cal C}]$. Consider then a reference equilibrium configuration $\mathcal{C}_{\rm eq}$, which is taken from the equilibrium Gibbs distribution, ${\cal P} ({\cal C}_{\rm eq}) = e^{- \beta {\cal H} (\mathcal{C_{\rm eq}})}/Z$,  and denote the overlap at point ${\bf x}$ between a configuration $\cal C$ and the reference configuration as $Q_{\bf x} (\mathcal{C},\mathcal{C}_{\rm eq}) = \delta_{{\cal C},{\cal C}_{\rm eq}}$. (For a liquid formed by $N$ particles one needs to introduce a smoothing  function $f(y)$ with a short range of the order of the cage size corresponding to the typical extent of the vibrational motions, {\it i.e.}, $Q_{\bf x} [\hat \r(\mathcal C),\hat \r(\mathcal C_{\rm eq})] = \int {\rm d} {\bf y} f(y) [ \hat{\r} ({\bf x} + {\bf y}\vert \mathcal C) \hat{\r}({\bf x} - {\bf y}\vert \mathcal C_{\rm eq}) - \r^2]$, where $\hat\rho(\mathbf x\vert \mathcal C)$ is the microscopic density at point $\bf x$ for a configuration $\mathcal{C}$ of the liquid and $\r = N/V$ is the average density: see the companion paper.\cite{paper2}) 

One can now define an overlap field $p({\bf x})$ and introduce an effective Hamiltonian or action for this field, which is the large-deviation functional describing the probability to observe a certain profile of the overlap field:
\begin{equation}
\label{eq:Spx} 
{\cal S}[p |\mathcal{C}_{\rm eq} ] \equiv 
- \ln \big( {\cal P} [p|\mathcal{C}_{\rm eq}] \big) = - \ln \left[ \frac{1}{Z} \sum_{\mathcal{C}} e^{- \beta {\cal H} (\mathcal{C})}
\, \delta \left [ p({\bf x}) - Q_{\bf x} (\mathcal{C},\mathcal{C}_{\rm eq}) \right ] \right]\,.
\end{equation}
For a uniform overlap $p({\bf x}) = p$ and in the mean-field limit, the action $\overline{{\cal S}[p |\mathcal{C}_{\rm eq}]}$, averaged over all different choices of the equilibrium configuration, becomes the Franz-Parisi potential $V(p)$.\cite{franzparisi-potential,franzparisi-potential2,franzparisi-potential3} This object encodes in a compact way the properties of the complex free-energy landscape of glassy systems. Between $T_d$ and $T_K$ it exhibits an absolute minimum in $p \simeq 0$ and a secondary minimum in $p_\star$, which corresponds to the overlap for a typical metastable state sampled at equilibrium (also called non-ergodicity parameter and related to the Debye-Waller factor) and whose height difference with the value at the stable minimum corresponds to the configurational entropy $s_c$. Qualitatively, $V(p)$ exactly behaves as the Landau free-energy of a $\f^4$ scalar field theory in the presence of a negative external magnetic field $H$, 
which describes a first-order transition from a negative to a positive magnetization at $H=0$. Pushing the analogy with magnetic systems a step further,\cite{Stevenson}  the overlap order parameter $p$ plays the role of the magnetization $m$, the configurational entropy $s_c$ is the counterpart of (minus) the external magnetic field $H$. Furthermore, a surface tension-like term $\g$, related to the height of the barrier between the two minima and called amorphous surface tension in the context of supercooled liquids, is proportional to the ferromagnetic  coupling $J$. Finally, the thermodynamic glass transition at $T_K$, at which the two minima in $p=0$ and $p = p_\star$ have the same free-energy,  corresponds to the first-order transition in $H=0$, whereas the dynamical glass transition at $T_d$ coincides with the spinodal of the positively magnetized state.

\begin{figure}
\includegraphics[scale=0.3]{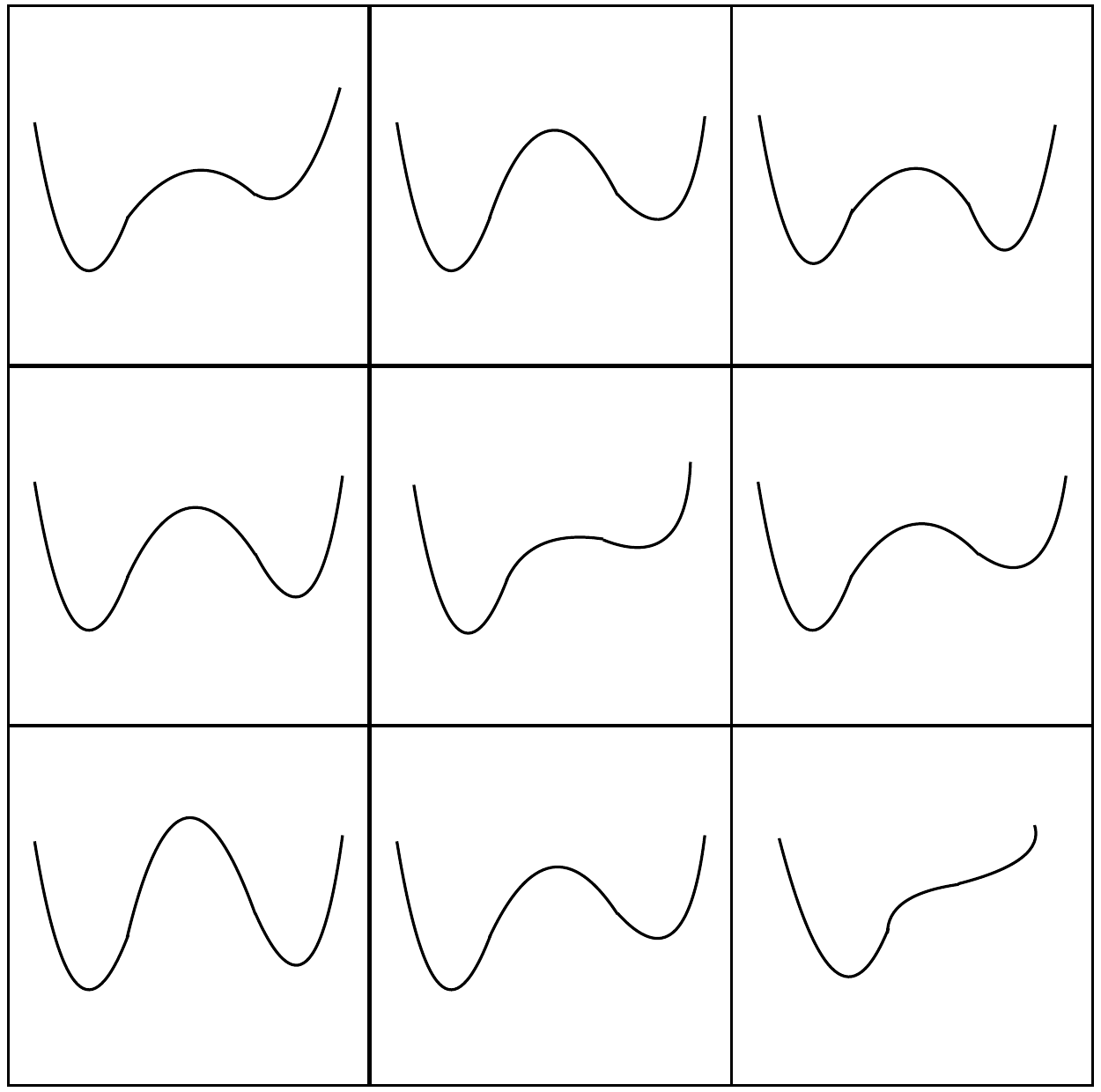}
\caption{Sketch of the distribution of local Franz-Parisi potentials when the system is divided in cubic boxes larger than the size of the particles but smaller than the point-to-set correlation length. This illustrates the local fluctuations of the configurational entropy and of the surface tension.}
\label{fig:boxes-RFIM}
\end{figure}

With this in mind, one can now argue that the reference equilibrium configuration acts as a quenched disorder. Indeed, although after averaging over $\mathcal{C}_{\rm eq}$  the global Franz-Parisi potential becomes independent of the reference configuration, its local properties still depend on the choice of $\mathcal{C}_{\rm eq}$ due to  the density fluctuations of the reference configuration. Imagine coarse-graining the system on a scale that is larger than the microscopic scale (\ie the size of the particles or the lattice spacing)  but smaller than the point-to-set length (above which metastability and configurational entropy are no longer  well defined in finite dimensions\cite{BB,tarzia-cammarota-etal}) by dividing the sample in cubic boxes  as sketched in Fig.~\ref{fig:boxes-RFIM} and computing the Franz-Parisi potential in each of these finite-size boxes considered as independent one from another. One would then observe fluctuations of the shape of the Franz-Parisi potential from one box to another, due to the local density fluctuations of the  reference configuration (see Fig.~\ref{fig:boxes-RFIM} for an illustration). This results in fluctuations of the height of the secondary minimum (\ie of the configurational entropy $s_c$ akin to a magnetic field) and of the height of the barrier between the two minima (\ie of the surface tension $\g$ akin to a  ferromagnetic coupling) among boxes. At a coarse-grained level, this naturally leads to a description in terms of a scalar $\f^4$ effective theory with quenched disorder in the form of a random field and a random bond. (Note that the above procedure can also be operationally implemented in computer simulations of glass-forming liquid models: This will be further discussed in the companion paper.\cite{paper2})

If one now tries to extend these phenomenological arguments to the vicinity of the putative thermodynamic glass (RFOT) transition at $T_K$,  one realizes that the different boxes may become strongly correlated due to the presence of a diverging  point-to-set correlation length.\cite{BB} Showing that the description based 
on an effective random-field + random-bond $2$-state  Ising-like theory is not jeopardised by these long-range correlations is a challenge. 
The results of the present paper and of its companion one\cite{paper2} suggest that this description continues to hold, but that the presence of a diverging point-to-set correlation length generically leads to the emergence of additional features such as multi-body interactions.

\section{Effective theory for mean-field models of structural glasses} 
\label{sec:mean_field}

In the first part of this paper we focus on two archetypal mean-field models for the glass formation, {\it i.e.}, the Random Energy Model (REM)\cite{derrida} and the fully connected version of its Kac-like generalization to a finite number ($2^M$) of states\cite{KREM} (called in the following the $2^M$-KREM), and we work out, essentially exactly, the effective $2$-state random Hamiltonian that describes the fluctuations of the overlap with a reference equilibrium configuration.

\subsection{Replicas and expansion in cumulants}
\label{sec:stage}

Because of the reference configuration, $\mathcal{C}_{\rm eq}$, the action ${\cal S} [p({\bf x})|\mathcal{C}_{\rm eq}]$, introduced in Eq.~(\ref{eq:Spx}), describes a generic scalar field theory in the presence of quenched disorder. In order to analyze it and understand in more detail what kind of disorder is generated by $\mathcal{C}_{\rm eq}$ one can study the cumulants of ${\cal S}$ by considering replicas of the original system. As known in the context of the critical behavior of the RFIM,\cite{Tarjus-Tissier1,Tarjus-Tissier2} $\exp (-{\cal S}_{\rm rep} [\{ p_a ({\bf x}) \}] ) = \overline{\exp (-\sum_{a=1}^n {\cal S} [p_a ({\bf x})|\mathcal{C}_{\rm eq}])}$ generates the cumulants of the action ${\cal S}[p({\bf x})|\mathcal{C}_{\rm eq}]$ through an expansion in increasing number of free replica sums:
\beq 
\label{eq:FRS}
{\cal S}_{\rm rep} [\{ p_a ({\bf x}) \}] = \sum_{a=1}^n {\cal S}_1 [p_a ({\bf x})] - \frac{1}{2} \sum_{a,b=1}^n {\cal S}_2 [p_a ({\bf x}), p_b ({\bf x})]
+ \frac{1}{3!} \sum_{a,b,c=1}^n {\cal S}_3 [p_a ({\bf x}), p_b ({\bf x}), p_c ({\bf x})] + \ldots \, ,
\eeq
where ${\cal S}_l [p_1, \ldots, p_l]$ is the $l$-th cumulant of ${\cal S}[p({\bf x})|\mathcal{C}_{\rm eq}]$, 
\beq \label{eq:S1S2_def}
\beal
{\cal S}_1 [p ({\bf x})] & = \overline{{\cal S}[p({\bf x})|\mathcal{C}_{\rm eq}]} \, , \\
{\cal S}_2 [p_1 ({\bf x}), p_2 ({\bf x})] & = \overline{{\cal S}[p_1({\bf x})|\mathcal{C}_{\rm eq}] {\cal S}[p_2({\bf x})|\mathcal{C}_{\rm eq}]}
- \overline{{\cal S}[p_1({\bf x})|\mathcal{C}_{\rm eq}]} \, \overline{{\cal S}[p_2({\bf x})|\mathcal{C}_{\rm eq}]} \, ,
\eal
\eeq
etc.

To compute the replicated action ${\cal S}_{\rm rep} [\{ p_a ({\bf x}) \}]$ one needs to perform the average over the reference configuration. This is achieved by introducing $n+1$ replicas, which will be identified by Greek letters $\a = 0 ,1, \ldots, n$ (whereas Roman letters $a = 1, \ldots, n$ will still be used for replicas from $1$ to $n$ only): 
\beq \label{eq:Srep}
\beal
e^{-{\cal S}_{\rm rep} [\{ p_a ({\bf x}) \}]}  & = \frac{1}{Z} \sum_{\mathcal{C}_{\rm eq},\mathcal{C}_a}
e^{- \b {\cal H} (\mathcal{C}_{\rm eq}) } \, 
e^{- \b \sum_{a=1}^n {\cal H} (\mathcal{C}_a) } \prod_{a=1}^n \delta \left [ p_a({\bf x}) - Q_{\bf x} (\mathcal{C}_a,\mathcal{C}_{\rm eq}) \right ] \\
& = \frac{1}{Z} \sum_{\mathcal{C}_\a}
e^{- \b \sum_{a=1}^n {\cal H} (\mathcal{C}_\a) } \prod_{a=1}^n \delta \left [ p_a({\bf x}) - Q_{\bf x} (\mathcal{C}_a,\mathcal{C}_{\rm eq}) \right ] \, ,
\eal
\eeq
where $\mathcal{C}_0 \equiv \mathcal{C}_{\rm eq}$.

\subsection{An illustrative toy model: the REM}
\label{sec:REM}

The REM is the simplest mean-field system displaying a thermodynamic glass transition (RFOT), and it therefore represents a natural first benchmark for our analysis.  Interestingly, the effective theory  for this model can be worked out without resorting to any approximation.

The REM, which was introduced by Derrida,\cite{derrida} is a disordered spin model defined as follows: The energies $E({\mathcal C})$ of the configurations $\mathcal C=\{S_1,\dots,S_N\}$, where $S_i=\pm 1$, are independent and identically distributed (i.i.d.) random variables sampled from a Gaussian distribution with zero mean and variance equal to $N/2$.  The REM displays a RFOT at $T_K=1/(2\sqrt{\ln 2})$. Below this temperature the system freezes in the lowest available states whereas above it the configurational entropy is positive. (The temperature $T_d$ of the dynamical glass transition is infinite in this model.)   The main reason for the simplicity of the REM lies in the fact that ``states'' and ``configurations'' coincide, {\it i.e.}, the intra-state entropy is zero. In consequence, the overlap $p$ takes only the values one and zero, which respectively correspond to the two replicas being in the same or in different states/configurations.

Since this model has explicit quenched disorder (the random energies), we have to perform an additional average over the distribution of this disorder. Hence,  Eq.~(\ref{eq:Srep}) now reads     
\begin{equation}
\begin{aligned}
 \label{eq_replica_action-rem}
e^{-{\cal S}_{\rm rep}
[\{p_a\}]} &=\frac{1}{\overline{Z}}\overline{\sum_{{\mathcal C}_0,\dots,{\mathcal C}_n}\exp\bigg(-\beta \sum_{\alpha} E({\mathcal C}_\alpha)\bigg)\prod_{a=1}^n \delta_{p_a,
\delta_{{\mathcal C}_0,{\mathcal C}_a}}}
=\frac{1}{\overline{Z}}\sum_{{\mathcal C}_0,\dots,{\mathcal C}_n}\exp\bigg(\frac{\beta^2 N}{4} 
\sum_{\alpha,\beta}\delta_{{\mathcal C}_\alpha,{\mathcal C}_\beta} 
\bigg)\prod_{a=1}^n \delta_{p_a,\delta_{{\mathcal C}_0,{\mathcal C}_a}} \, ,
\end{aligned}
\end{equation}
where we have used the ``annealed approximation'', exact only for $T\ge T_K$,\cite{quenched_barrier} in which one does not need to introduce another set of replicas to handle the average over the random energies. By using the fact that two replicas having an overlap one with the reference configuration also have a mutual overlap equal to one, it is straightforward to obtain that the term in the argument of the exponential in Eq.~(\ref{eq_replica_action-rem}) can be rewritten as
\[
\sum_{\alpha,\beta}\delta_{{\mathcal C}_\alpha,{\mathcal C}_\beta}=
1 + 2 \sum_{a=1}^n \delta_{{\mathcal C}_a,{\mathcal C}_0} + \sum_{a,b}\delta_{{\mathcal C}_a,{\mathcal C}_b}
= 1 + 2 \sum_{a=1}^n p_a + \sum_{a,b} p_a p_b + \sum_{a,b}^\star \delta_{{\mathcal C}_a,{\mathcal C}_b} =
\Big(1+\sum_a p_a \Big)^2+\sum_{a,b}^\star \delta_{{\mathcal C}_a,{\mathcal C}_b} \, ,
\]
where $\sum_{a,b}^\star$ denotes a sum that only runs over replicas $a$ and $b$ having a zero overlap with the reference configuration. Using this result we can rewrite the partition function as
\begin{equation}
\label{eq_replica_action-rem2}
e^{-{\cal S}_{\rm rep} 
[\{p_a\}]} \propto \sum_{{\mathcal C}_0}  e^{\frac{\beta^2 N}{4} \left(1+\sum_a p_a \right)^2 } 
\sum_{{\mathcal C}_1,\dots,{\mathcal C}_n}^\star \exp\bigg(\frac{\beta^2 N}{4} 
\sum_{a,b}^\star \delta_{{\mathcal C}_a,{\mathcal C}_b}
\bigg) \, ,
\end{equation}
where again the star in the sum over configurations means that one has to sum only over replicas having a zero overlap with the reference configuration (and thus only over $2^N-1$ configurations different from ${\mathcal C}_0$).  The term between parentheses is nothing else than the replicated partition function of a REM with a number of replicas equal to $n'=n-\sum_a p_a$ (and $2^N-1$ available configurations). one can use the replica method to compute it. Since $0\le n' \le n\rightarrow 0$ and we consider $T\ge T_K$, the replicated partition function appearing in Eq.~(\ref{eq_replica_action-rem2}) can be obtained as\cite{KREM}
\[
\sum_{{\mathcal C}_1,\dots,{\mathcal C}_n}^\star \exp\bigg(\frac{\beta^2 N}{4} 
\sum_{a,b}^\star \delta_{{\mathcal C}_a,{\mathcal C}_b}
\bigg)
\approx \exp \Big [ \Big( n-\sum_a p_a \Big) \Big(N\ln 2 +\frac{\beta^2 N}{4} \Big) \Big ] \, ,
\]
which coincides with the annealed approximation. After collecting all these results together we find the following expression for the replicated action:
\begin{equation}\label{eq:Srep_REM}
\begin{aligned}
{\cal S}_{\rm rep} [\{p_a\}]
=- (n + 1) \Big(N\ln 2 +\frac{\beta^2N}{4}\Big)+ \Big(N\ln 2 -\frac{\beta^2 N}{4} \Big) \sum_a p_a 
- \frac{\beta^2 N}{4} \Big(\sum_a p_a\Big)^2 \, .
\end{aligned}
\end{equation}
This can be directly interpreted as the replicated action for a two-state variable $p=0,1$, or equivalently as the replicated action for the Hamiltonian of an Ising variable $\sigma=2p-1$ coupled to a Gaussian random magnetic field,
\begin{equation}
\label{eq_REM}
\beta{\cal H}_{\rm eff} = {\rm cst} - (\mu + \d \m) p=  {\cal S}_0- (H + \d h) \sigma \, ,
\end{equation}
with $H = \mu/2 = -N (4 \ln 2 - \beta^2)/8$, $\overline{\d h} = 0$, $\overline{\d h^2} = \overline{\d \m^2}/4 = N\beta^2/8$, and ${\cal S}_0={\rm cst}-\delta h$, which is the result already given in Eq.~(\ref{eq:Heff_REM}) of the Introduction. (By enforcing $H=0$ one recovers the value of the critical temperature of the REM, as expected.) 
Thus, the statistics of the fluctuations of the overlap with a reference configuration, which is described by ${\cal S} [p({\bf x})|\mathcal{C}_{\rm eq}]$, is the same as that of an Ising (or discrete global overlap) variable that is subjected to a field with an average value of the order of $N$, favoring the $\sigma=-1$ (zero overlap) state and vanishing at the transition temperature, and with fluctuations of the order of $\sqrt N$. (The role of the random energy ${\cal S}_0$ is to ensure that the variance of the effective Hamiltonian is equal to zero in the zero-overlap, or $\sigma=-1$,  state.)

In conclusion we have found that the theory describing the overlap fluctuations of the REM is a $0$-dimensional RFIM. We did not attempt to generalize the computation for temperatures below $T_K$ but from known results on the REM, we expect to find a disordered action corresponding to an Ising-like variable coupled to an external random field whose typical strength is of the order of one. The analysis performed in this section shows that without any approximation the RFIM naturally emerges in the study of glassy systems for temperatures below $T_d$.

\subsection{Effective theory for the fully connected $2^M$-KREM}
\label{sec:REMFC}

In this section we consider a nontrivial, but still exactly solvable, generalization of the REM introduced for the first time in Ref.~[\onlinecite{KREM}], the $2^M$-KREM on a fully connected lattice. We apply the strategy outlined in Sec.~\ref{sec:stage} to obtain the (quasi) exact effective theory that describes the statistics of
the fluctuations of the overlap profile with an equilibrium reference configuration between $T_d$ and $T_K$. This theory will turn out to be given by a fully connected random field + random bond Ising model with multi-body interactions [see Eq.~(\ref{eq:Heff_REMFC})].

The model is defined as follows: Given $N$ sites, on each site $i$ there are $2^M$ configurations, $\mathcal C_i=\{1,\cdots,2^M\}$, and on each link $(i,j)$ we define i.i.d. Gaussian random energies $E_{ij}=E(\mathcal C_i,\mathcal C_j)$ with $\overline {E_{i j} (\mathcal{C}_i, \mathcal{C}_j)} = 0$ and $\overline {E_{i j} (\mathcal{C}_i, \mathcal{C}_j) E_{i j} (\mathcal{C}_i^\prime, \mathcal{C}_j^\prime)} = M \delta_{\mathcal{C}_i,\mathcal{C}_i^\prime} \delta_{\mathcal{C}_j,\mathcal{C}_j^\prime}$.
The Hamiltonian of the model is simply given by
\begin{displaymath}
\mathcal{H} = \frac{1}{2 \sqrt{N}} \sum_{i \neq j} E_{i j} (\mathcal{C}_i, \mathcal{C}_j) \,.
\end{displaymath}
The more standard mean-field result corresponds to $M \to \infty$ ($M$ plays the same role as $N$ in the simple REM discussed above). The model can be solved exactly by using replicas, as shown in Appendix~\ref{app:REMFC-1RSB}, and the thermodynamic glass transition (RFOT) taking place at $T_K$ can be obtained within a $1$-step replica-symmetry-breaking ($1$-RSB) ansatz.\cite{1RSB}

To construct the effective theory we consider $n+1$ replicas of the system and compute the replicated action for a fixed overlap configuration $\{ p_a^i \}$ of the replicas $a=1,\ldots,n$ with a given equilibrium reference configuration $\{ {\cal C}_i^0 \}$. Note that $p_a^i = 1$ only if $\mathcal{C}_i^a = \mathcal{C}_i^0$ and is zero otherwise. As already mentioned, we consider the temperature range $T_d \le T \le T_K$, where we can use the annealed approximation to perform the average over the random energies. (We stress that it is crucial on the other hand that the average over the quenched disorder represented by the reference configuration is performed exactly.) The replicated action then reads
\begin{equation} 
\label{eq:Srep-REMFC}
\begin{split}
e^{-\mathcal{S}_{\rm rep}[\{p_a^i\}]} &= \frac{1}{\overline{Z}} \overline{\sum_{\{ \mathcal{C}_i^\alpha \}} 
e^{- \frac{\beta}{2 \sqrt{N}} \sum_{ i \neq j,\alpha} 
E_{i j} (\mathcal{C}_i^\alpha, \mathcal{C}_j^\alpha) }
\prod_{a,i} 
\delta_{p_a^i,\delta_{\mathcal{C}_i^0,\mathcal{C}_i^a}}}
= \frac{1}{\overline{Z}}
\sum_{\{ \mathcal{C}_i^\alpha \}}  
e^{\frac{\beta^2 M}{8N} \sum_{ i \neq j } \sum_{\alpha,\beta=0}^n
\delta_{\mathcal{C}_i^\alpha,\mathcal{C}_i^\beta} \delta_{\mathcal{C}_j^\alpha,\mathcal{C}_j^\beta} }
\prod_{a,i} 
\delta_{p_a^i,\delta_{\mathcal{C}_i^0,\mathcal{C}_i^a}} \, .
\end{split}
\end{equation}
The Kronecker $\delta$'s in the exponential of the above expression can be rewritten in terms of the overlap variables as
\begin{equation} \label{eq:sum}
\sum_{\alpha,\beta=0}^n 
\delta_{\mathcal{C}_i^\alpha,\mathcal{C}_i^\beta} \delta_{\mathcal{C}_j^\alpha,\mathcal{C}_j^\beta}
= 1 + n + 2 \sum_{a=1}^n p_a^i p_a^j + \sum_{a\neq b=1}^n \delta_{\mathcal{C}_i^a,\mathcal{C}_i^b} \delta_{\mathcal{C}_j^a,\mathcal{C}_j^b}  \, .
\end{equation}
We note that if $\mathcal{C}_i^a = \mathcal{C}_i^0$ and $\mathcal{C}_i^b = \mathcal{C}_i^0$ ({\it i.e.}, $p_a^i = p_b^i = 1$), then $\mathcal{C}_i^b = \mathcal{C}_i^a$. Similarly, if $\mathcal{C}_i^a = \mathcal{C}_i^0$ and $\mathcal{C}_i^b \neq \mathcal{C}_i^0$ ({\it i.e.}, $p_a^i = 1$ and $p_b^i = 0$), then $\mathcal{C}_i^b \neq \mathcal{C}_i^a$. The same is true, of course, if $\mathcal{C}_i^a \neq \mathcal{C}_i^0$ and $\mathcal{C}_i^b = \mathcal{C}_i^0$. The only undetermined case corresponds to $\mathcal{C}_i^a \neq \mathcal{C}_i^0$ and $\mathcal{C}_i^b \neq \mathcal{C}_i^0$.

As discussed above, the expansion of $\mathcal{S}_{\rm rep}[\{p_a^i\}]$ in an increasing number of unrestricted sums over replicas, Eq.~(\ref{eq:FRS}), generates the cumulants of the effective disordered Hamiltonian describing the fluctuations of the overlap with a reference configuration. Below, we compute the first and second cumulants of such an effective Hamiltonian, which correspond to the $1$-replica and $2$-replica components of the replicated action [see Eq.~(\ref{eq:S1S2_def})].

\subsubsection{First cumulant of the effective disordered Hamiltonian}
\label{sec:REMFC-first}

Let us first focus on the first cumulant $\mathcal S_1 [\{p^i\}]$. From Eq.~(\ref{eq:FRS}) one realizes that the simplest way of computing it is to set all replica fields equal, $p_a^i =p^i$ $\forall \, a=1, \cdots,n$ and $\forall \, i$, keep only the term of order $n$ in the expression of $\mathcal S_{\rm rep}[\{p_a^i\}]$, and take the limit $n \to 0$  in the end, as in the standard replica trick.\cite{noi_Tc} After averaging over the random energies and the reference configuration, all the sites become equivalent and ${\cal S}_1 [\{p^i\}]$ can only be a function of $c = (1/N) \sum_i p^i$, which coincides with the global mean overlap with the reference configuration. Therefore we will use  ${\cal S}_1 (c)$ in place of ${\cal S}_1 [\{p^i\}]$ in what follows. This implies that scanning over all the possible configurations of the overlap profile $\{ p^i \}$ is the exact analogue of setting the overlap with the reference configuration for all replicas to be $1$ on the first $cN$ sites (\ie $p_a^i = 1$ for $i=1,\ldots,cN$ and $\forall a$) and $0$ on all the other $(1-c)N$ sites (\ie $p_a^i = 0$ for $i=cN+1,\ldots,N$ and $\forall a$).

Although this procedure resembles that of the random pinning,\cite{Cammarota_pinning,Cammarota1} it is different in that on the sites where $p^i=0$ the replicas {\it cannot} be in the same configuration as the reference one. It is also different from the computation of the Franz-Parisi potential\cite{franzparisi-potential,franzparisi-potential2,franzparisi-potential3} discussed in Sec.~\ref{sec:self-induced}. 
The reason is that in the latter case one sums over all configurations in which all replicas have the same fixed global overlap with the reference configurations ($\sum_i p_a^i = cN$~$\forall a$) but with different replicas having in  general different overlap profiles ($p_a^i \neq p_b^i$), whereas in the present procedure one restricts the sum to configurations in which {\it all} the $n$ replicas are constrained to have the {\it same} specific overlap profile with $\{ {\cal C}_i^0 \}$ ($p_a^i = p^i$~$\forall a$, such that $\sum_i p^i = cN$).



Our basic idea is to evaluate ${\cal S}_1 (c)$ by expanding it for small $c$ around $c=0$, ${\cal S}_1 (c) = \sum_q {\cal S}_1^{(q)} (0) c^q/q!$, and, since any power of $c$ can be re-expressed in terms of effective interactions among the $p^i$'s, \eg $c^q = (1/N^{q-1}) \sum_{i_1, \ldots, i_q} p^{i_1} \cdots p^{i_q}$, one can re-interpret the expansion of the $1$-replica component ${\cal S}_1 (c)$ as the average part of an effective diosrdered Hamiltonian with multi-body interactions of the form
\begin{equation} \label{eq:Heff_expansion}
{\cal S}_1 (c)= {\rm cst} - \mu \sum_i p^i - \frac{w_2}{2 N} \sum_{i \neq j} p^i p^j - \frac{w_3}{3! N^2} \sum_{i,j,k\neq} p^i p^j p^k - 
\frac{w_4}{4! N^3} \sum_{i , j,k,l \neq} p^i p^j p^k p^l + \ldots \, ,
\end{equation}
with $\mu = - {\cal S}_1^{(1)} (0)$, and $w_q =  - {\cal S}_1^{(q)} (0)$.
Since we are interested here in obtaining the effective Hamiltonian for $T \geq T_K$, the typical equilibrium configurations of the overlap are expected to have a small number of sites where  $p^i=1$. We thus look for an effective Hamiltonian which is accurate for small $c$ and we expect that the first few coefficients of the expansion of ${\cal S}_1(c)$ are sufficient to reproduce its behavior correctly. (Again, reconstructing the behavior of $\mathcal{S}_1 (c)$ at large $c$ is less important, since configurations with $c$ close to $1$ are very rare for $T > T_K$.) Note that the strategy presented here to compute ${\cal S}_1 (c)$ can  in principle be straightforwardly applied to any mean-field model in the same ``universality class'' as the REM and the $2^M$-KREM, with a complex free-energy landscape and a thermodynamic glass transition (RFOT), possibly with some minor and model-dependent modifications. 

For the chosen overlap profile, Eq.~(\ref{eq:sum}) becomes (after dropping sub-extensive terms)
\[
\beal
\sum_{i \neq j} \Big[ 1 + n + 2 n p^i p^j + \sum_{a\neq b} \delta_{\mathcal{C}_i^a,\mathcal{C}_i^b} \delta_{\mathcal{C}_j^a,\mathcal{C}_j^b}
\Big]  
\approx & (1 + n) N^2 + n (n + 1) c^2 N^2 + 2 c N \sum_{a \neq b} \sum_i^\star \delta_{\mathcal{C}_i^a,\mathcal{C}_i^b} 
+ \sum_{a \neq b} \Big( \sum_i^\star \delta_{\mathcal{C}_i^a,\mathcal{C}_i^b} \Big)^2 \, ,
\eal
\]
where the sum $\sum_i^\star$ represents the sum over the sites $i = cN + 1, \ldots, N$ where $p^i=0$. Inserting this expression into Eq.~(\ref{eq:Srep-REMFC}) yields
\[
e^{-n {\cal S}_1 (c)} = \frac{e^{\frac{N \beta^2 M}{8} \left [ 1 + n + n (n+1) c^2 \right ]}}{\overline{Z}} 
\sum_{\{ \mathcal{C}_i^\alpha \}} e^{\frac{N \beta^2 M}{8} \left[ 2 c \sum_{a \neq b} \frac{1}{N} 
\sum_i^\star \delta_{\mathcal{C}_i^a,\mathcal{C}_i^b} + \sum_{a \neq b} \left( \frac{1}{N} 
\sum_i^\star \delta_{\mathcal{C}_i^a,\mathcal{C}_i^b} \right)^2 \right]}
\prod_{a,i} 
\delta_{p^i,\delta_{\mathcal{C}_i^0,\mathcal{C}_i^a}}
\,. 
\]
On the first $cN$ sites, $p^i=1$ and, accordingly, ${\cal C}_i^a = {\cal C}_i^0$ for all $a$. Since the sum over the reference configuration ${\cal C}_i^0$ simply gives $2^{NM}$, we thus obtain
\[
e^{-n {\cal S}_1 (c)} = \frac{e^{N M \left \{ \ln 2 + \frac{\beta^2}{8} \left [ 1 + n + n (n+1) c^2 \right ] \right \} }}{\overline{Z}}
\sum_{\{ \mathcal{C}_i^a \}_\star } e^{\frac{\beta^2 M}{4} \sum_{a < b} \left [ 2 c \sum_i^\star \delta_{\mathcal{C}_i^a,\mathcal{C}_i^b}
+ \frac{1}{N} \left( \sum_i^\star \delta_{\mathcal{C}_i^a,\mathcal{C}_i^b} \right)^2 \right] } \, ,
\]
where the trace $\sum_{\{ \mathcal{C}_i^a \}_\star }$ represents the sum over all the $2^M - 1$ configurations $\mathcal{C}_i^a$ different from the reference one on the $(c-1)N$ sites where $p^i=0$.

One can now introduce the overlaps $q_{ab}$ by performing $n(n-1)/2$ Hubbard-Stratonovich transformations (all the details of the calculations are reported in Appendix~\ref{app:REMFC_Eff_1}). We posit a replica-symmetric (RS) ansatz for the overlap matrix, $q_{ab} = q_0$, which is, again, expected to be justified for $T \ge T_K$ and for small $c$, as it is for instance for the Franz-Parisi potential \cite{franzparisi-potential3} where the RS ansatz is appropriate for small and large enough values of the overlap ({\it i.e.}, of $c$), whereas one needs to use a 1-RSB ansatz for intermediate values of $c$, around the barrier. This is quite clear on physical grounds, as increasing $c$ effectively  reduces the configurational entropy that is accessible to the constrained system, thereby inducing a $1$-RSB glass transition at moderately large values of the overlap, whereas for larger values of $c$ the system is constrained to be in same configuration as the reference one, which corresponds to a RS ansatz. In the following we will be mostly interested in the lowest-order coefficients of the expansion of ${\cal S}_1 (c)$ in powers of $c$, which give the dominant effective interactions in Eq. (\ref{eq:Heff_expansion}). The values of these coefficients can therefore be computed by means of a RS ansatz. Furthermore, our goal here is not to obtain the complete analytic expression of ${\cal S}_1 (c)$, but just to illustrate the general strategy allowing us to compute it. Computing ${\cal S}_1 (c)$ within a 1-RSB ansatz is certainly doable, but would just make the calculation much more cumbersome and involved without changing the general picture.

In the thermodynamic limit ($N \to \infty$), by using the saddle-point method, we finally derive (see Appendix~\ref{app:REMFC_Eff_1} for details)
\begin{equation} \label{eq:S1c}
\frac{{\cal S}_1 (c)}{N} = -\frac{\beta^2 M}{8} (1 + c^2) - \frac{\beta^2 M q_0^2}{8} + \frac{(1 - c) \beta^2 M (q_0 + c)}{4}
- (1 - c) 
\, \overline{\ln \Bigg[ \sum_{{\cal C}}^\star e^{\sqrt{\frac{\beta^2 M (q_0 + c)}{2}} z_{{\cal C}}} \Bigg]} \, ,
\end{equation}
where the overlap $q_0$ satisfies the following self-consistent equation:
\begin{equation} \label{eq:q0}
q_0 = (1 - c) \left \{ 1 - \sqrt{\frac{2}{\beta^2 M (q_0 + c)}} 
\overline{\left(\frac{\sum_{{\cal C}}^\star z_{{\cal C}} \, e^{\sqrt{\frac{\beta^2 M (q_0 + c)}{2}} z_{{\cal C}}}}
{\sum_{{\cal C}}^\star e^{\sqrt{\frac{\beta^2 M (q_0 + c)}{2}} z_{{\cal C}}}}\right)} \right \} \, .
\end{equation}
The averages of the form $\overline{[ f(\vec{z}_{\cal C})]}$ appearing in the above expressions are defined over a Gaussian measure,
$\overline{[ f(\vec{z}_{\cal C})]} \equiv \int \prod_{{\cal C}}^\star \left[ \frac{{\rm d}z_{\cal C}}{\sqrt{2 \pi}} \, e^{- z_{\cal C}^2/2} \right] f(\vec{z}_{\cal C})$, with $\vec{z}_{\cal C}$ being a $(2^M-1)$-dimensional vector. The solution of such a saddle-point equation can be expanded in powers of $c$ as $q_0 \approx q_{0,0} +  q_{0,1}c + q_{0,2} c^2 + \ldots$, which, when inserted back into Eq.~(\ref{eq:S1c}), allows one to obtain the exact coefficients of the expansion of ${\cal S}_1 (c)$.

We illustrate the output by providing explicit analytical expressions for the chemical potential $\mu$ and the coupling constants $w_q$ appearing in Eq.~(\ref{eq:Heff_expansion}) in large-$M$ limit. In this limit, $q_0$ is of order of $1/2^M$. In consequence, expanding the exponentials in Eqs.~(\ref{eq:S1c}) and (\ref{eq:q0}) up to the eighth order in $\sqrt{M \beta^2 (q_0 + c)/2}$ gives Eq.~(\ref{eq:q0-Kac}) of Appendix~\ref{app:REMFC_Eff_1}, which, when inserted into Eq.~(\ref{eq:S1c}), leads to the following expressions of $\mu$ and of $w_q$ up to the fourth order: 
\beq \label{eq:Kn}
\begin{aligned}
{\rm cst} & =-\frac{M \beta^2}{8} - M \ln 2\, \\
\mu & = - M \ln 2 - \frac{M \beta^2}{2^{M+2}} \, \\
w_2 & = \frac{M \beta^2}{4} - \frac{M \beta^2(M \beta^2 - 4)}{2^{M+3}} \, \\
w_3 & = - \frac{M^2 \beta^4(M \beta^2 - 6)}{2^{M+4}} \, , \\
w_4 & = - \frac{M^3 \beta^6(M \beta^2 - 8)}{2^{M+5}} \, .
\end{aligned}
\eeq
When $M \to \infty$ all the interactions beyond the pairwise one vanish (as in the REM case: see above). Note also that all the coupling constants $w_q$ seem to decrease as the temperature is increased, in agreement with physical intuition.

\begin{figure}
\centering
\includegraphics[width=0.54\textwidth]{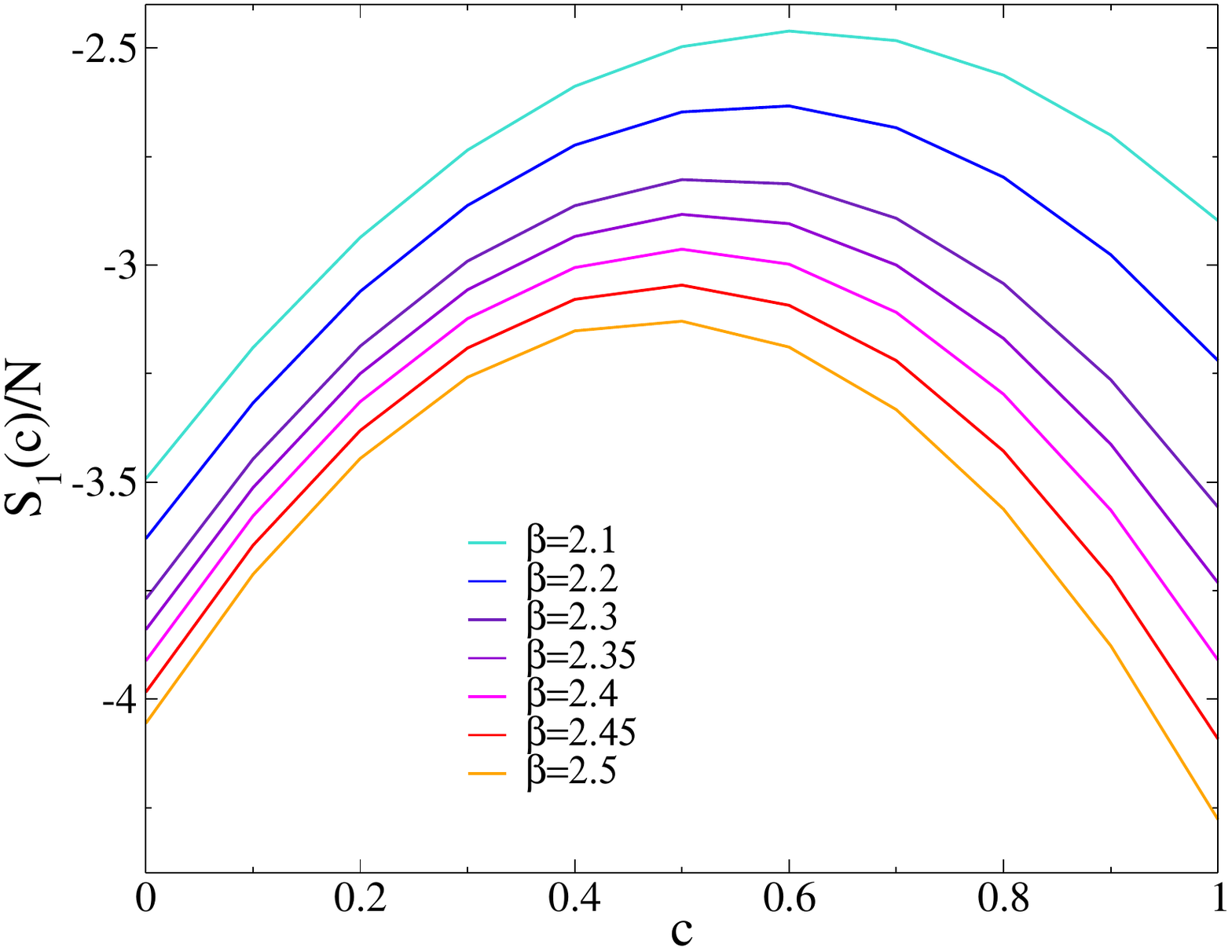}
\caption{$1$-replica part ${\cal S}_1(c)$ of the replicated effective action as a function of the overlap $c$ with a reference equilibrium configuration for the fully connected $2^M$-KREM with $M=3$ [Eqs.~(\ref{eq:S1c}) and~(\ref{eq:q0})]. Several values of the temperature,  $T \geq T_K\approx 0.4$ ({\it i.e.}, $\beta\leq \beta_K\approx 2.5$), are shown.}
\label{fig:S1-quasiexact}
\end{figure}

For finite values of $M$, another strategy to determine the coefficients of the effective Hamiltonian consists instead in solving Eq.~(\ref{eq:q0}) numerically for several values of $c$ and $\beta$, inserting the result into Eq.~(\ref{eq:S1c}), and fitting the function ${\cal S}_1(c)$ so obtained by a polynomial function of $c$. This yields the values of the coefficients $\mu$ and $w_q$ of the effective Hamiltonian, as well as their temperature dependence. The procedure is illustrated in Fig.~\ref{fig:S1-quasiexact} for $M=3$, where ${\cal S}_1 (c)$ is plotted for several values of $\beta$ for $T \gtrsim T_K$. Fitting these curves with polynomials of $c$ of degree $4$ provides a numerical estimate of $\mu$, $w_2$, $w_3$, and $w_4$ for different temperatures. 

\subsubsection{Second cumulant of the effective disordered Hamiltonian}
\label{sec:REMFC-second}

We now turn to the computation of the second cumulant, {\it i.e.}, the $2$-replica part of the replicated action defined in Eq.~(\ref{eq:Srep-REMFC}). To do this we divide the $n$ constrained replicas into two groups of $n_1$ and $n_2$ replicas respectively. The most generic overlap profile can be obtained by dividing the sites into four groups, denoted $(1)$, $(2)$, $(12)$, and $(0)$, such that on the $c_1 N$ sites belonging to the group $(1)$ $p_1^i=1$ and $p_2^i=0$, on the $c_2 N$ sites belonging to the group $(2)$ $p_1^i=0$ and $p_2^i=1$, on the $c_{12} N$ sites belonging to the group $(12)$ $p_1^i = p_2^i = 1$, and on the $c_0 N$ sites belonging to the group $(0)$ $p_1^i = p_2^i = 0$ (with $c_0 = 1-c_1-c_2-c_{12}$). The second cumulant can be computed by keeping only the terms of order $n_1 n_2$ in the expression of the replicated action, and taking the limit $n_1 , n_2 \to 0$ [see Eq.~(\ref{eq:FRS}) and Ref.~[\onlinecite{noi_Tc}]]. The key observation is again that the second cumulant can only be a function of $c_1$, $c_2$, and $c_{12}$, and thus of the global overlaps [see Eq.~(\ref{eq:c1c2c12})].

Our general strategy is similar to that used for the first cumulant:

\textbf{1.} We first perform the standard Hubbard-Stratonovich transformations that allow us to decouple the sites of the fully connected lattice
via three overlap matrices, which correspond to the overlaps between two different constrained replicas belonging to the first group, $q_{ab}^{[1]}$, to the second group, $q_{ab}^{[2]}$,  and to the two different groups, $q_{ab}^{[12]}$.

\textbf{2.} We then posit a RS ansatz for the overlaps, $q_{ab}^{[1]}=q_1$ $\forall a \neq b$, $q_{ab}^{[2]} = q_2$ $\forall a \neq b$, and $q_{ab}^{[12]} = q_{12}$ $\forall a,b$, which is expected to be justified for $T \ge T_K$ and for small enough $c_1$, $c_2$, and $c_{12}$ , and we perform the trace over the configurations.

\textbf{3.} In the thermodynamic limit the saddle-point method yields the expression of the replicated free energy in terms of the overlaps [Eqs.~(\ref{eq:A2}) and~(\ref{eq:A_n1n2})], with the latter obeying three self-consistent equations [Eqs.~(\ref{eq:q1}) and (\ref{eq:q12})].

\textbf{4.} In order to compute the terms of order $n_1 n_2$ of the replicated action, which yields the second cumulant of the effective Hamiltonian, we expand the RS overlaps as
\begin{equation} \label{eq:q_exp}
\beal
q_1 & \approx q_1^{[0,0]} + n_2 q_1^{[0,1]} + O(n_1,n_2^2,n_1 n_2) \, , \\
q_2 & \approx q_2^{[0,0]} + n_1 q_2^{[1,0]} + O(n_2,n_1^2,n_1 n_2) \, , \\
q_{12} & \approx q_{12}^{[0,0]} + O(n_1,n_2) \, .
\eal
\end{equation}
Inserting this expansion into the saddle-point equations, Eqs.~(\ref{eq:q1}) and (\ref{eq:q12}), allows us to obtain $q_1^{[0,0]}$, $q_1^{[0,1]}$, $q_2^{[0,0]}$, $q_2^{[1,0]}$, and $q_{12}^{[0,0]}$ which, when inserted into the expression of the replicated action, Eq.~(\ref{eq:A_n1n2}), finally leads to the second cumulant as a function of the concentrations $c_1$, $c_2$, and $c_{12}$.

\textbf{5.} In practice, we are interested in the expansion of ${\cal S}_2 (c_1,c_2,c_{12})$ only  up to the second order in the concentrations of the different kinds  of sites, $c_1$, $c_2$ and $c_{12}$,  which corresponds to the most relevant random terms. In fact, it is easy to realize that any power of the concentrations can be re-expressed as effective random terms in the expression of the second cumulant through
\begin{equation} \label{eq:c1c2c12}
c_1 
= \frac{1}{N} \sum_i p_1^i (1 - p_2^i) \, , 
\qquad
c_2 
= \frac{1}{N} \sum_i p_2^i (1 - p_1^i) \, , 
\qquad
c_{12} 
= \frac{1}{N} \sum_i p_1^i p_2^i \, .
\end{equation}

The calculations, although conceptually simple, are long and tedious. In consequence, for the sake of the clarity of the presentation, we give the explicit expression of the second cumulant in the large-$M$ limit only, up to the leading order in $1/2^M$, and we defer all the details to Appendix~\ref{app:REMFC_Eff_2}:
\beq 
\label{eq:S2_2MKREM}
\beal
{\cal S}_2[\{p_1^i,p_2^i \}] & \approx \frac{M \beta^2}{4} \bigg( 1 + \frac{M \beta^2 + 4}{2^{M+1}}
\bigg) \frac{1}{N} \sum_{i \neq j} p_1^i p_2^i p_1^j p_2^j + \frac{M \beta^2}{2^{M+1}} \bigg[1 - \frac{1}{N} \sum_i \big( p_1^i + p_2^i  \big) \bigg] 
\sum_j p_1^j p_2^j   \, .
\eal
\eeq

\subsubsection{The effective disordered Hamiltonian}

Following Sec.~\ref{sec:stage}, the expansion in replica sums of the replicated action is equivalent to the expansion in cumulants of a disordered Hamiltonian ${\mathcal H}_{\rm eff}[p] \equiv {\cal S} [p({\bf x})|\mathcal{C}_{\rm eq}]$, 
with the identification $\mathcal S_1 [p_a]=\overline{\beta \mathcal H_{\rm eff}[p_a]}$, $\mathcal S_2 [p_a,p_b]=\overline{\beta \mathcal H_{\rm eff}[p_a]\beta \mathcal H_{\rm eff}[p_b]}-\overline{\beta \mathcal H_{\rm eff}[p_a]}\,\overline{\beta \mathcal H_{\rm eff}[p_b]}$, etc., where the overline now denotes an average over the effective quenched disorder. 
The form of the effective disordered Hamiltonian that is able to reproduce the $1$- and $2$-replica parts derived above reads 
\[
\beta {{\cal H}_{\rm eff}} = {\rm cst} -  \sum_i (\mu+ \delta \mu_i) p^i - \frac{1}{2} \sum_{i \neq j}  \Big ( \frac{w_2}{N} + \frac{\delta w_{2,ij}}{\sqrt{N}} \Big) p^i p^j 
- \frac{w_3}{3! N^2} \sum_{i,j,k\neq} p^i p^j p^k -
\frac{w_4}{4! N^3} \sum_{i , j,k,l \neq} p^i p^j p^k p^l + \ldots \, ,
\]
where the chemical potential $\mu$ and the couplings $w_q$ are given in Eq.~(\ref{eq:Kn}), and the covariances of the quenched random variables $\delta \mu_i$, $\delta w_{2,ij}$ have to be chosen in order to reproduce the expression of ${\cal S}_2$ in Eq.~(\ref{eq:S2_2MKREM}), namely,
\begin{equation}
\begin{aligned}
\overline{\delta \mu_i \delta \mu_j} & = \frac{M \beta^2}{2^{M+1}} \delta_{ij} \, ,\\
\overline{\delta w_{2,ij} \delta w_{2,kl}} & = \frac{M \beta^2}{2} ( \delta_{ik} \delta_{jl} + \delta_{il} \delta_{jk} ) \, , \\
\overline{\delta \mu_i \delta w_{2,jk}} & = - \frac{M \beta^2}{2^M} \frac{\delta_{ij} + \delta_{ik}}{\sqrt{N}} \, ,
\end{aligned}
\end{equation}
and $\overline{\delta \mu_i} = \overline{\delta w_{2,ij}} = 0$.

Going as before from overlap variables to spin variables, $p^i = (1 + \sigma^i)/2$, one finally obtains the effective random-field + random-bond fully connected Ising model $\beta {{\cal H}_{\rm eff}}[\sigma]$ given in Eq.~(\ref{eq:Heff_REMFC}).  
The explicit expressions of the external field and the coupling constants for $M \gg 1$ are
\begin{equation} \label{eq:HJ2J3}
\begin{aligned}
H & = \left [\frac{\mu}{2} + \frac{w_2}{4} + \frac{w_3}{16} + \frac{w_4}{96} +
\ldots\right ] \approx \frac{1}{2} \Big[ \frac{M \beta^2}{8} - M \ln 2
- \frac{M^2 \beta^4}{3 \cdot 2^{M+9}} \big(24 + 4 M \beta^2 + M^2 \beta^4 \big) +\ldots \Big ] \, \\
J_2& =  \frac{w_2}{4} + \frac{w_3}{16} + \frac{w_4}{32} + \ldots \approx  \frac{1}{16} \Big[ M \beta^2 
+ \frac{M \beta^2}{2^{M+6}}
\big( 128 + 16 M \beta^2 - M^2 \beta^4 \big)+\cdots  \Big] \, \\
J_3 & = \frac{w_3}{8} + \frac{w_4}{16} + \ldots \approx \frac{M^2 \beta^4(24 + 4 M \beta^2 - M^2 \beta^4+\cdots)}{2^{M+9}} \, , \\
J_4 & = \frac{w_4}{16} + \ldots \approx \frac{M^3 \beta^6(8 - M \beta^2+\cdots )}{2^{M+9}} \, ,
\end{aligned}
\end{equation}
and the random fields and random couplings are characterized by
\beq \label{eq:HJ2J3-var}
\beal
\overline{\delta h_i \delta h_j} & \approx \frac{M \beta^2}{32} \big( 1 - 2^{2-M} \big ) \delta_{ij} + \frac{M \beta^2}{32 N} \big( 1 - 2^{3-M} \big ) \, , \\
\overline{\delta J_{2,ij} \delta J_{2,kl}} & \approx \frac{M \beta^2}{32} ( \delta_{ik} \delta_{jl} + \delta_{il} \delta_{jk} ) \, , \\
\overline{\delta h_i \delta J_{2,jk}} & \approx \frac{M \beta^2}{32} \big (1 + 2^{2-M} \big) \frac{\delta_{ij} + \delta_{ik}}{\sqrt{N}} \, ,
\eal
\eeq 
and $\overline{\delta h_i} = \overline{\delta J_{2,ij}} = 0$.

Note that compared to the REM studied in the preceding section, the introduction of a finite number of states, $2^M$, leads to additional multi-body interactions and additional random terms. The additional multi-body interactions vanish in the limit $M \to \infty$. In this limit the cumulants of the random variables simplify, but one nonetheless remains with both random fields and random bonds. This difference between $2^M$-KREM and REM stems from the fact that the random energies are defined on the links of the lattice in the former and on the sites in the latter.

\subsubsection{Comparison with the exact solution}

\begin{figure}
\centering
\includegraphics[width=0.54\textwidth]{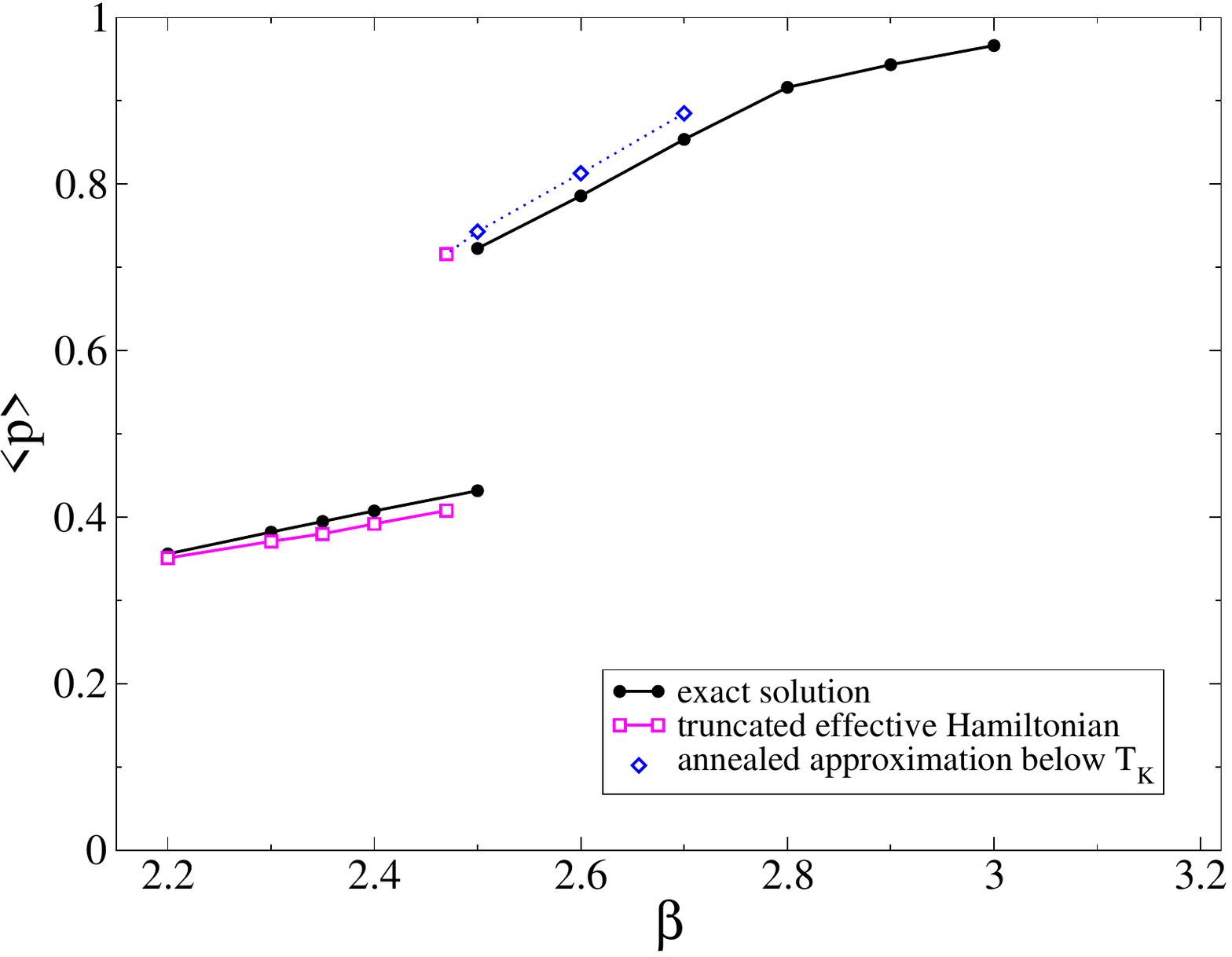}
\caption{Comparison between the (truncated) effective theory and the exact result for the fully connected $2^M$-KREM with $M = 3$: Mean overlap $\langle p\rangle$ with a reference equilibrium configuration versus the inverse temperature $\beta$. The black curve (circles) represents the exact solution [Eqs.~(\ref{eq:SPRS}) and~(\ref{eq:SPm})], whereas the red curve (squares) corresponds to the prediction of the effective random-field + random-bond Ising Hamiltonian, with the coupling constants of the multi-body interactions (up to $4$-body) numerically extracted  from Fig.~\ref{fig:S1-quasiexact} as discussed in the text and the covariances of the random variables given in Eq.~(\ref{eq:HJ2J3-var}). The counterpart of the overlaps in the disordered Ising model is the magnetization $(1 + m)/2$, where $m = \overline{\langle \sigma_i \rangle}$ [see Appendix~\ref{app:REMFC-RFIMI} and Eqs.~(\ref{eq:REMFC-solution})]. The effective disordered theory has been derived by using the annealed approximation to average over the random energies, which is in principle justified only above $T_K$ and at $T_K$ (which  includes the jump of $\langle p\rangle$). The blue curve (diamonds) shows the prediction of the effective theory below $T_K$ with an additional approximation, {\it i.e.}, when the coupling constants and the covariances are obtained by continuing the results obtained via the annealed approximation in the low-temperature phase.}
\label{fig:overlap}
\end{figure}

The main result of this section is that the effective theory describing the probability distribution of the thermal fluctuations of the overlap with an  equilibrium configuration for the fully connected $2^M$-KREM corresponds to a random-field + random-bond fully connected Ising model with multi-body interactions. In practice, we need to truncate the number of multi-body interactions ({\it e.g.}, up to $4$-body terms, as done above) and truncate as well the expansion in cumulants of the random variables ({\it e.g.}, keeping only the second cumulants). In order to test the quantitative accuracy of the truncated effective theory we have computed its prediction for the mean overlap $\langle p\rangle$ with the reference configuration [which in the Ising model is simply related to the magnetization $m$ through $\langle p\rangle=(1+m)/2$] as a function of temperature and compared it to the exact result derive in Appendix~\ref{app:REMFC-1RSB}. (To further simplify the computation we have dropped the cross-correlations between the random fields and the random bonds and we have neglected the off-diagonal term of the random-field distribution: Details can be found in Appendix~\ref{app:REMFC-RFIMI}.) The comparison for the case $M=3$ is displayed in Fig.~\ref{fig:overlap}. One can see that  there is a very good agreement, which thus shows that neglecting higher-order interactions and cumulants is not only qualitatively and but also quantitatively justified.

Note that, since we have used the annealed approximation to perform the average over the random energies in the derivation of the effective theory, our results are
in principle only valid on the high-temperature side of the RFOT, $T\ge T_K$, including the RFOT itself. Our procedure could be extended to the low-temperature side by performing a quenched average over the random energies. Although standard, the computation becomes long and tedious in this case. Moreover, the aim of our analysis is not to provide an accurate determination of the numerical values of the various parameters entering the effective disordered Hamiltonian deep in the low-temperature glass phase; it is instead to show how and why such an effective description emerges in a general, transparent, and robust way, and to analyze the properties of the transition point. For these reasons the results derived from the effective disordered Ising Hamiltonian in the region $T<T_K$ ($T_K$ corresponds to a first-order transition with a jump of the magnetization in the effective description) are simply obtained with approximate coupling constants and covariances determined  through the annealed approximation.The agreement with the exact solution is nonetheless quite good.

For finite $M$  the expansion of the first and second cumulants  ${\cal S}_1 (c)$ and ${\cal S}_2 (c_1,c_2,c_{12})$ in powers of $c$,  $c_1$, $c_2$, and $c_{12}$ generates multi-body interactions and random terms to all orders (higher-order cumulants are present as well). Still, the dominant term that controls the transition  in the effective theory is expected to be given by the competition between the ferromagnetic tendency of the interactions and the fluctuations of the random fields,  and therefore to display RFIM-like behavior. The effect of the random bonds could become important if $\sqrt{\overline{\delta J_{ij}^2}} \gg J_2$, as this could generate a spin-glass-like behavior, as advocated in Ref.~[\onlinecite{Moore}]. Yet, this possibility is excluded for the structural-glass model considered here for any $M>1$, since $\sqrt{\overline{\delta J_{ij}^2}}/J_2 \sim 1/\sqrt{M}$.)

We stress a key difference between the $2^M$-KREM with $M$ finite and the REM. Despite being a fully connected mean-field model, the former is indeed such that the fluctuations associated with the effective disorder coming from the reference configuration give a contribution to the thermodynamics that is of the same order as that of the the average part. The difference of course disappears when $M\to \infty$, and disorder-related fluctuations are then subdominant as they scale as $\sqrt{M}$ for the $2^M$-KREM and $\sqrt N$ for the REM whereas the average contributions scale as $M$ and $N$, respectively.

\section{Discussion}
\label{discussion}

\subsection{Beyond mean-field: Illustration of the difficulties} 
\label{sec:beyondMF}

We now want to illustrate some of the difficulties that one encounters when trying to generalize the procedure developed in the preceding sections to finite-dimensional glass-forming systems close to the putative thermodynamic glass transition. To make the presentation more concrete we focus on a paradigmatic spin model of structural glass, the spherical $p$-spin model and consider its Kac extension that allows one to go one step beyond the conventional mean-field limit by taking into account spatially heterogeneous solutions.\cite{Franz_Kac,Franz_instantons} 

The model is defined by the following Hamiltonian:
\begin{equation}
\begin{aligned} 
\label{eq_kac_pspin}
\mathcal H = -\sum_{i_1\cdots i_p \in \Lambda} J_{i_1\cdots i_p} \sigma_{i_1}\cdots \sigma_{i_p}
\end{aligned}
\end{equation}
where  the spin variables satisfy the spherical constraint and the coupling constants  $J_{i_1\cdots i_p}$ are i.i.d. Gaussian variables with variance
\begin{equation}
\begin{aligned} 
\label{eq_kac_variance}
\overline{J_{i_1\cdots i_p}^2}=\frac 1{r_0^{pd}}\sum_{k \in \Lambda} \psi\left(\frac{\vert k-i_1 \vert}{r_0} \right )\cdots \psi\left(\frac{\vert k-i_p \vert}{r_0} \right ) \, ,
\end{aligned}
\end{equation}
where the function $\psi(r)$ is well-behaved and decays on a scale of order O($1$). The Kac limit consists in considering the limit of a large interaction range, $r_0\to \infty$.

The action $\mathcal S[\{p_a,q_{ab}\}]$ for the overlaps $\{p_a\}$ between the reference and the ``constrained'' replicas  and for the overlaps $\{q_{ab}\}$ among the constrained replicas can then be obtained by standard methods and is given at large distance by\cite{Franz_Kac}
\begin{equation}
\begin{aligned}
 \label{eq_replica_action_pspin}
\mathcal S[\{q_{\alpha\beta}\}]\approx r_0^d \int_x \bigg \{\frac{\beta c}{2} \sum_{\alpha\beta \neq} (\partial_x q_{\alpha\beta}(x))^2
-\frac{\beta^2}{4} \sum_{\alpha\beta \neq} q_{\alpha\beta}(x)^p  - \frac 12 \mathrm{Tr} \log[\boldmath{I} 
+\boldmath{U}(\{q_{\alpha\beta}(x)\})]
\bigg \} \, ,
\end{aligned}
\end{equation}
where $\boldmath{I}$ is the identity and $\boldmath{U}$ an $(n+1)\times (n+1)$ matrix with all diagonal elements equal to $0$, $U_{0a}=q_{0a}=p_a$, $U_{a0}=q_{a0}=p_a$, and $U_{ab}=q_{ab}$ for $a \neq b$. As before, Greek letters are used for the $n+1$ copies of the original system, including the reference equilibrium configuration $\alpha=0$, whereas Latin ones are reserved for the $n$ replicas other than the reference one. The action for the overlap fields $p_a$ is then obtained by integrating out the overlaps $q_{ab}$,
\begin{equation}
\begin{aligned}
 \label{eq_replica_action2}
e^{-S_{\rm rep}[\{p_a\}]} \propto \int  \prod_{ab \neq} \mathcal D q_{ab} \, e^{- \mathcal S[\{p_a,q_{ab}\}]} \,.
\end{aligned}
\end{equation}
where, in the Kac limit, the integral over the $q_{ab}$'s can be performed via a saddle-point calculation, {\it i.e.},
\begin{equation}
S_{\rm rep}[\{p_a\}]={\rm cst} + \mathcal S[\{p_a,q_{ab}^*\}] \, ,
\end{equation}
with
\begin{equation}
\frac{\partial S[\{p_a,q_{ab}\}]}{\partial q_{ab}(x)}\bigg \vert^*=0 \, .
\end{equation}
Above the thermodynamic glass transition at $T_K$, the first cumulant is then given by (see Ref.~[\onlinecite{noi_Tc}] for a detailed derivation)
\begin{equation}
\begin{aligned} 
\label{eq:cumulant1}
{\cal S}_1[p_1] = r_0^d \int_x \Big \{ \beta c (\partial_x p_1 (x))^2 
	- \frac{\beta c}{2} (\partial_x q^* (x))^2 
	- \frac{\beta^2}{4} \left [ 2 p_1(x)^p - q^*(x)^p \right] + \frac{p_1(x)^2 - q^*(x)}{2[1 - q^*(x)]} - \frac{1}{2} \log [1 - q^*(x)] \Big \},
\end{aligned}
\end{equation}
where $q^*(x)$ satisfies the following saddle-point equation:
\begin{equation} 
\label{eq:sp1a}
\beta c\partial_x^2q^*(x) +\frac{\beta^2 p}{4} q^*(x)^{p-1} = \frac{1}{2} \, \frac{q^*(x)-p_1(x)^2}{[1-q^*(x)]^2}\, .
\end{equation}
Unfortunately, the same letter $p$ is used to denote the number of spins involved in the interactions in Eq.~(\ref{eq_kac_pspin}) (this is a widespread notation that it would awkward to change) and the overlap with the reference configuration. To avoid too much confusion, our convention is that $p$ always comes with a replica index, $p_a$, when it refers to an overlap.

The solution of Eq.~(\ref{eq:sp1a}) is in general far from being trivial. When the thermodynamics of the system is dominated by specific uniform profiles $p_a(x)=p_a$ and smooth variations around them ({\it e.g.}, in the Kac limit, $r_0 \to \infty$), one can first solve Eq.~(\ref{eq:sp1a}) for uniform $p_1$ and $q$ and then consider the first nonzero gradient corrections about the uniform solution. More generally, for a large but finite $r_0$, gradient expansions may not be enough and one needs in principle to take into account the contribution coming from nonuniform profiles $p_a(x)$. However, even assuming that one can use the saddle-point equation~(\ref{eq:sp1a}) to compute the integral over the $q_{ab}$'s, one immediately sees that solving Eq.~(\ref{eq:sp1a}) for a generic nonuniform $p_1(x)$ is an impossible task. Approximations are therefore required.

A chief obstacle to devising a simple approximation for solving Eq.~(\ref{eq:sp1a}) comes from the presence of specific spatial correlations that arise in the form of point-to-set correlations. At high enough temperature, but still below the dynamical transition temperature $T_d$, these correlations are short-ranged and one can therefore proceed as in the Kac limit by considering uniform configurations and smooth variations that can be described by an expansion in spatial derivatives. On the contrary when one approaches $T_K$ one anticipates long-ranged, possibly diverging, point-to-set correlations.
Consider then for instance a configuration of $p_1(x)$ that is zero almost everywhere except for a finite density of localized regions of space where it has a high value corresponding to the glassy metastable minimum. One expects that $q(x)$ will also be equal to the metastable high-overlap value in the these localized regions. What is then the value of $q(x)$ in the regions where $p_1(x)=0$? If there were no point-to-set correlations, $q(x)$ would be zero. However, as soon as the typical distance between high-overlap regions becomes less than the point-to-set correlation length, the constraint due to the high-overlap regions will suddenly force the overlap $q(x)$ in the rest of the system to take a nontrivial value distinct from zero. At the mean-field $T_K$ (in the Kac limit) the point-to-set correlation length is infinite\cite{Franz_instantons} and even a very dilute concentration of localized high-overlap regions will induce a nontrivial finite features in the profile $q(x)$ everywhere. This example shows how nonperturbative changes of $q(x)$ can be generated by minute changes in the profile of $p_1(x)$, leading as a result to intrinsically nonlocal and long-ranged contributions to ${\cal S}_1[p_1]$. This is the essence of the difficulty associated with deriving an effective theory for the $p_a$'s specifically when the system is near or below the mean-field thermodynamic glass transition. We will address this issue in the following paper.\cite{paper2}

\subsection{Specific features of the effective disorder} 
\label{sec:features_disorder}

Another point which is worth discussing concerns the properties of the effective disorder found when mapping the statistics of the fluctuations of the overlaps $\{p_a\}$ onto a disordered Ising model. To illustrate this we again consider the Kac spherical $p$-spin model and we focus on the second cumulant ($2$-replica action) ${\cal S}_2[p_1,p_2]$. In the limit $r_0\to \infty$, we find after some manipulations (see Ref.~[\onlinecite{noi_Tc}] for a detailed derivation) that
\begin{equation} 
\label{eq:cumulant2}
\begin{aligned} 
{\cal S}_2[p_1,p_2] = \int_x \bigg \{- \beta c (\partial_x q_{12}(x))^2 + \frac{\beta c}{2} \big[ \big( \partial_x  q_1^{[0,1]}(x)  \big)^2 + \big( \partial_x q_2^{[1,0]} (x) \big)^2 \big] + \frac{\beta^2}{2} [q_{12}(x)]^p - \frac{[q_{12}(x) - p_1(x) p_2(x)]^2}{2 [1 - q_1(x)] [1 - q_2(x)]} \bigg \} \, ,
\end{aligned}
\end{equation}
where $q_{12}(p_1,p_2)$,  $q_1(p_1)$, $q_2(p_2)$, $q_1^{[0,1]} (p_1,p_2)$, and $q_2^{[1,0]} (p_1,p_2)$ satisfy the following saddle-point equations:
\begin{equation}
\label{eq:sp12}
\begin{aligned}
2 \beta c \partial_x^2 q_{12}(x) + \frac{\beta^2 p}{2} [q_{12}]^{p-1} & = \frac{q_{12}- p_1 p_2}{(1 - q_1) (1 - q_2)} \,, \\
\beta c \partial_x^2 q_a(x) + \frac{\beta^2 p}{4} q_a^{p-1} & = \frac{1}{2} \, \frac{q_a-p_a^2}{[1-q_a]^2} \, , \\
	- \beta c \partial_x^2 q_a^{[\cdot,\cdot]_a} + \bigg[ p(p-1) \frac{\beta^2}{4} q_a^{p-2} - \frac{1 - 2 p_a^2 + q_a}{2 (1 - q_a)^3} \bigg] q_a^{[\cdot,\cdot]_a} &= - \frac{(q_{12} - p_a p_b )^2}{2 (1 - q_a)^2 (1-q_b)} \, ,
\end{aligned}
\end{equation}
with $a=1,2$, $[\cdot,\cdot]_1 = [0,1]$ and $[\cdot,\cdot]_2 = [1,0]$.

As already discussed, in the Kac limit on which we focus here, one only needs to consider uniform profiles $p_a(x)=p_a$ and smooth variations around them. One can then first solve the saddle-point equations in Eq.~(\ref{eq:sp12}) for uniform $p_1$ and $p_2$ and then calculate the first nonzero gradient correction about the uniform solution. In the following we focus on the local part of the second cumulant $\mathcal S_2$ which is obtained by considering uniform overlaps, as it is sufficient to illustrate our point.

\begin{figure}
\centering
\includegraphics[width=0.54\textwidth]{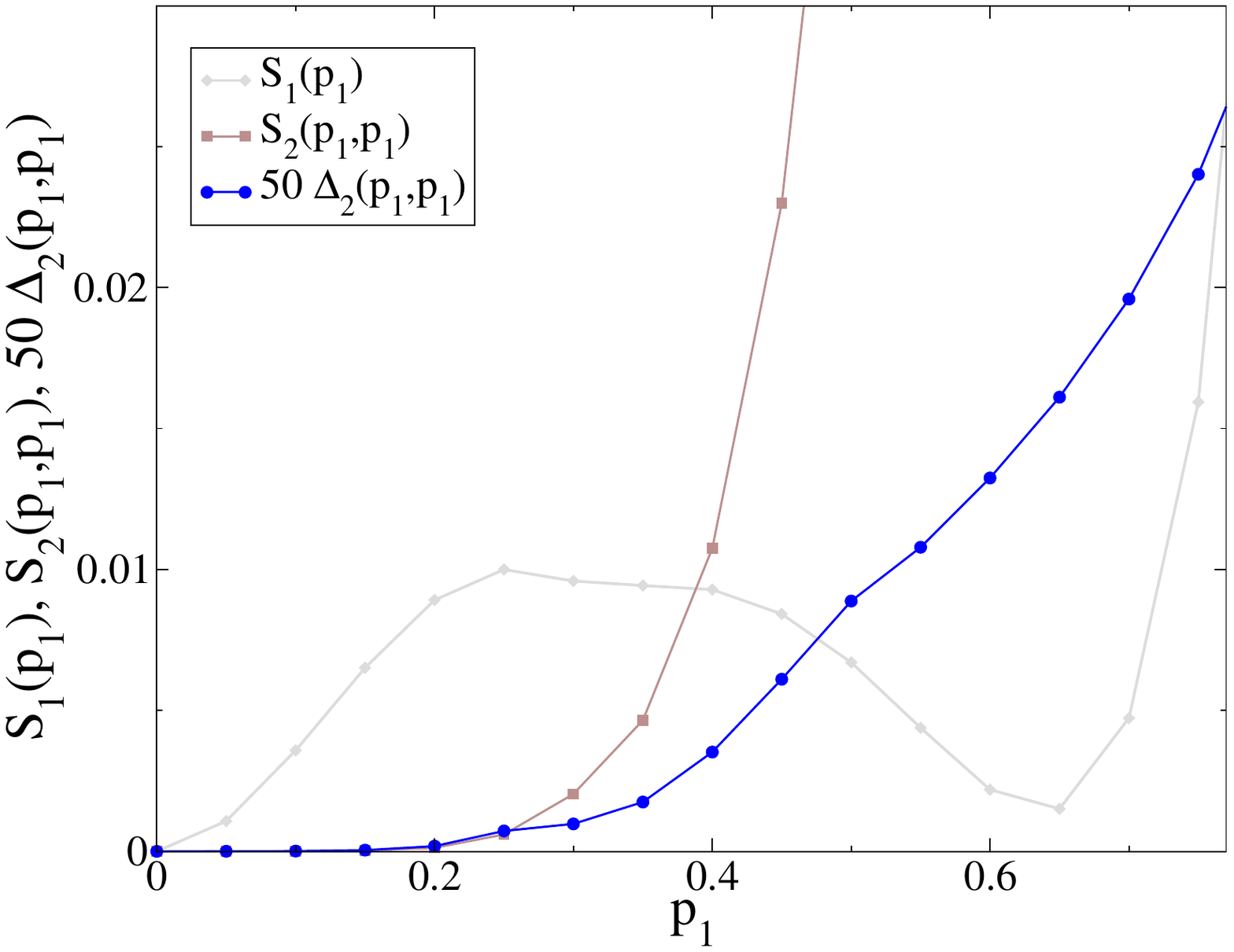}
\caption{Local part of the variance of the effective random field, $\Delta_2(p_1,p_2)=\partial_{p_1}\partial_{p_2}S_2(p_1,p_2)$, and of the $2$-replica action, $S_2(p_1,p_2)/N$, for equal arguments $p_1 = p_2$ as a function of $p_1$ (blue curve, circles) for the $p$-spin model in the Kac limit with $p=3$ and $\beta = 1.7$ (such that $\beta_d < \beta < \beta_K$). $\Delta_2$ is zero in the liquid minimum ($p_1=0$), grows as $p_1$ is increased, so that it is strictly positive at the metastable glassy minimum. ${\cal S}_2 (p_1,p_1)$ behaves similarly. We also display the local part of the first cumulant ${\cal S}_1 (p_1)/N$ given in Eq.~(\ref{eq:cumulant1}), to show the position of the secondary minimum around $p_\star \approx 0.65$: grey curve and diamonds (note that the region around the barrier is not properly described by the present RS solution but this is irrelevant for our illustrative purpose).  $\Delta_2$ has been multiplied by $50$ to plot the three curves on the same scale.}
\label{fig:Delta2}
\end{figure}

The local part of the functional, $\mathcal S_2(p_1,p_2)/N$,  is simply related to the second cumulant of a (delta correlated in space) random potential. Its second derivative $\Delta_2(p_1,p_2)=\partial_{p_1}\partial_{p_2}[\mathcal S_2(p_1,p_2)/N]$ then represents the second cumulant of the derivative of the random potential, {\it i.e.}, the variance of an effective random force or random source conjugate to the overlap field. (After passing to the magnetic representation in terms of the magnetizations $m_a=2p_a-1$, $\Delta_2$ represents, up to a factor $4$, the variance of the effective random field.) It is instructive to analyze the shape of $\mathcal S_2(p_1,p_2)/N$ and $\Delta_2(p_1,p_2)$. Note first that ${\cal S}_2=0$ when either $p_1=0$ or $p_2=0$, which implies that the effective disorder vanishes in the liquid phase, as one could anticipate on the basis of intuitive arguments. For the Kac spherical $p$-spin model one also has $\Delta_2(p_1,p_2)=0$ when either $p_1=0$ or $p_2=0$. For illustration, we plot $\Delta_2(p_1,p_2)$ when $p_1=p_2$  in Fig.~\ref{fig:Delta2}. It is zero in $p_1=0$, as announced, and grows as $p_1$ increases to be strictly positive at the value $p_\star$ of the glassy metastable minimum. 

The property that the variance of the effective random field $\Delta_2(p_1,p_2)=0$ when $p_1=0$ or $p_2=0$ is not true, however, for the REM: From Eq.~(\ref{eq:Srep_REM}), one instead obtains a constant variance $\Delta_2 (p_1,p_2) = \beta^2 N /4$. In fact having a zero or nonzero value of $\Delta_2$ in the liquid minimum is related to the details of the microscopic description. It is indeed easy to see that a disordered model with quenched disorder in the couplings only, as the Kac spherical $p$-spin model, leads to an effective theory with $\Delta_2 = 0$ in the liquid, whereas if some local (on-site) quenched disorder is present at the microscopic level, the corresponding effective theory is characterized by ${\cal S}_2 = 0$ but $\Delta_2 > 0$ for either $p_1$ or $p_2$ equal to zero. There is actually a subtlety when considering glassy models in which the overlap degrees of freedom $p_a^i$ are hard binary variables taking only values $0$ and $1$, as in the case of the REM and of the $2^M$-KREM (but not of the Kac spherical $p$-spin model). What is, for instance, the counterpart for hard binary variables of a model with continuous variables and disorder in the interactions such that ${\cal S}_2(p_1,p_2)/N= (\Delta/4)p_1^2p_2^2$ ? Simply replacing the $p_a$'s by hard variables (on a lattice) with $p_a^2=p_a$ yields ${\cal S}_2(p_1,p_2)/N = (\Delta/4)p_1 p_2$. Whereas $\Delta_2(p_1,p_2)= \Delta p_1p_2$ in the former case and vanishes when either $p_1$ or $p_2$ equals zero, it is equal to a constant, $\Delta_2(p_1,p_2) = \Delta/4$, in the latter case.  For hard binary variables on a lattice the only way that the local part of the second cumulant ${\cal S}_2$ does not simply reduce to a term proportional to $p_1 p_2$ is that in the Ising (magnetic) representation, in addition to a local random-field term, random-bond disorder with cross-correlations with the random field is present. This is indeed the case for the $2^M$-KREM, see Eq.~(\ref{eq:S2_2MKREM}).\cite{footnote_S0}

\section{Concluding remarks} 
\label{sec:conclusions}

In this paper we have presented the derivation of a $2$-state random effective theory which describes the fluctuations of what is thought to be the relevant order parameter for glassy systems, \ie the overlap field with a reference equilibrium configuration, close to the putative thermodynamic glass transition temperature. We have focused on archetypal mean-field models for the glass transition, in particular the Random Energy Model (REM)\cite{derrida} and its version with a finite number of states\cite{KREM} on a fully connected lattice ($2^M$-KREM). The effective Hamiltonian for mean-field models can in principle be worked out without resorting to any approximation. We have shown that the statistics of the fluctuations of the overlap with a reference configuration for the REM are
described by an Ising variable $\s = \pm 1$ (corresponding to high and low overlap with the reference configuration) subjected to a random field whose average is of order $N$, and vanishes exactly at $T_K$, and whose fluctuations are of order $\sqrt{N}$. The effective theory for the fully connected $2^M$-KREM is richer, and it is given by a random-bond + random-field Ising model with multi-body interactions and higher-order random terms. 

We argue that the mapping is very general and should apply (possibly with some minor model-dependent adjustments) to any mean-field glassy model in the same ``universality class'' with a complex free-energy landscape appearing between an upper dynamical glass transition and a lower thermodynamical (RFOT) glass transition. In fact, we have shown that for mean-field models the first cumulant of the effective action can only be a function of the global overlap with the reference configuration, and its shape shown in Fig.~\ref{fig:Delta2} (although distinct) is similar to that of the Franz-Parisi potential.\cite{franzparisi-potential,franzparisi-potential2,franzparisi-potential3} In order to obtain the average part of the effective Hamiltonian one thus only needs to expand the first cumulant in powers of the global overlap and re-express the resulting terms as effective multi-body interactions. The same strategy can be straightforwardly generalized to the second (and higher) cumulant, which yields the fluctuations of the effective Hamiltonian. Although conceptually simple, the calculations are somehow long and tedious, except for the REM. In the case of the $2^M$-KREM we have derived for illustration the analytic expressions of the effective coupling constants and of the second cumulants of the random terms  when $M \gg 1$ and we have given a numerical recipe to compute them when $M$ is finite.

It is worth stressing that while the effective disorder is found to be subdominant in the $N\to \infty$ limit for the REM, the fluctuations associated with the effective disorder in the case of the fully connected $2^M$-KREM for finite $M$ are, in the thermodynamic limit, of the same order as the average contribution. Contrary to a naive expectation, the effective disorder is relevant for mean-field fully connected models: The choice of the reference configuration leads to site-dependent fluctuations of the ``local'' effective configurational entropy and of the ``local'' effective surface tension (as sketched in fig.~\ref{fig:boxes-RFIM}) which give a contribution to the thermodynamics at the leading order. One needs to integrate over the fluctuations of the local 
overlap. In order to do it properly, taking into account the effective disorder is crucial. 
This also highlights the main difference between $\mathcal{S}_1(c)$ and the Franz-Parisi potential: The former 
is the average of the action for the instantaneous local fluctuations of the overlap, whereas the latter 
is the thermodynamic potential associated to the global overlap.

We finally reiterate that the interest of deriving the effective theory for mean-field models of structural glasses which can of course be exactly solved by other means is twofold:

{\it (i)} It shows that an effective description in terms of a random-field + random-bond Ising model naturally emerges in a transparent and general way.

{\it (ii)} It justifies it on a quantitative basis. In fact, by allowing a direct comparison with exact results, it justifies the truncation of the effective theory to a limited number of multi-body interactions (typically, up to $4$-body), a limited order of cumulants (typically, up to the second one) and of quenched random terms (typically, random fields and random bonds). This serves as a guide for the investigation of finite-dimensional glass-forming systems. In a renormalization-group perspective, all higher-order terms will then anyhow be generated by the further renormalization of the effective theory to obtain the full solution of the thermodynamics of the overlaps. Once the effective theory is established, this final step can be achieved by using all powerful nonperturbative means at our disposal, such as large-scale numerical simulations\cite{RFIM_simulations} or the functional renormalization group.\cite{Tarjus-Tissier1,Tarjus-Tissier2}

{\it (iii)} It is relevant for real finite-dimensional liquids, where on a scale much larger than the microscopic length but still much smaller than the point-to-set correlation length the mean-field description is still expected to retain some validity. One can then construct the effective theory on this scale by taking the mean-field result as a starting point. This would lead to an effective Hamiltonian of the form given in Eq.~(\ref{eq:Heff_REMFC}). In order to derive the proper effective theory for finite-dimensional glass-formers near the putative thermodynamic glass transition, one then has to take into account the role of correlations on the scale of the (diverging) point-to-set length. As already stressed, this is the main issue to be solved to go beyond the mean-field description. We will tackle it in the following paper.\cite{paper2}

\begin{acknowledgments}
We acknowledge support from the ERC grant NPRGGLASS and the Simons Foundation grant on ``Cracking the Glass Problem''(No. 454935, GB).
\end{acknowledgments}

\appendix

\section{The fully connected Kac-like REM with a finite number of states: Exact solution, Effective theory, and Variational
approximation} 
\label{app:REMFC}

This appendix is devoted to the analysis of the Kac-like version of the REM with $2^M$ states (the $2^M$-KREM) on a fully connected lattice. The model, first introduced in Ref.~[\onlinecite{KREM}], is defined as follows: We consider $N$ sites and define a state variable $\mathcal{C}_i$ on each site $i$ which can can take $2^M$ possible values, $\mathcal{C}_i=1, \ldots, 2^M$. For each pair of sites $(i,j)$ we define the couplings $E_{ i j } (\mathcal{C}_i, \mathcal{C}_j)$, which are i.i.d.~Gaussian random variables such that $\overline {E_{i j} (\mathcal{C}_i, \mathcal{C}_j)} = 0$ and $\overline {E_{i j} (\mathcal{C}_i, \mathcal{C}_j) E_{i j} (\mathcal{C}_i^\prime, \mathcal{C}_j^\prime)} = M \delta_{\mathcal{C}_i,\mathcal{C}_i^\prime} \delta_{\mathcal{C}_j,\mathcal{C}_j^\prime}$. The Hamiltonian of the system is then given by $\mathcal{H} = 1/(2 \sqrt{N}) \sum_{i \neq j} E_{i j} (\mathcal{C}_i, \mathcal{C}_j)$. In the following, we begin by working out the exact solution of the model by using the standard replica approach and a $1$-RSB ansatz.

\subsection{Exact solution using the standard replica approach} 
\label{app:REMFC-1RSB}

In order to compute the free energy of the system we use the replica trick,
\begin{equation} \label{eq:Zn}
\begin{split}
\overline{Z^n} &= \overline{\sum_{\{ \mathcal{C}_i^\alpha \}} 
\exp \bigg( - \frac{\beta}{2\sqrt{N}} \sum_{i \neq j,\alpha} 
E_{i j} (\mathcal{C}_i^\alpha, \mathcal{C}_j^\alpha) \bigg) }
= e^{nNM \beta^2/8} \sum_{\{ \mathcal{C}_i^\alpha \}} \exp \bigg[ \frac{M \beta^2}{8N} \sum_{\alpha \neq \beta} \Big( \sum_i \delta_{\mathcal{C}_i^\alpha,\mathcal{C}_i^\beta} 
\Big)^2 \bigg] \, .
\end{split}
\end{equation}
A simple calculation shows that in the Kac limit ($M \to \infty$) the model has a (RFOT) glass transition at an inverse temperature $\beta_K=\sqrt{8 \ln 2}$ (see Sec.~\ref{sec:REM}). We expect that at finite $M$ the transition, if present, will be located at a lower temperature. After performing $n(n-1)/2$ Hubbard-Stratonovich trasformations Eq.~(\ref{eq:Zn}) can be rewritten as
\begin{displaymath} 
\overline{Z^n} = e^{nN M \beta^2 /8} \Big(\frac{N M \beta^2}{8 \pi} \Big)^{\!n(n-1)/2} \int \prod_{\alpha < \beta} \textrm{d} q_{\alpha \beta} \,
e^{-N A[q_{\alpha \beta}]} \, ,
\end{displaymath}
where
\begin{equation} \label{eq:A_qab}
 A[q_{\alpha \beta}] = \frac{M \beta^2}{8} \sum_{\alpha \neq \beta} q_{\alpha \beta}^2 - \ln Z_1 [q_{\alpha \beta}] \, ,
\qquad \textrm{with} \qquad
Z_1 [q_{\alpha \beta}] = \sum_{\{ \mathcal{C}^\alpha \}} e^{\frac{M \beta^2}{4} \sum_{\alpha \neq \beta} \delta_{\mathcal{C}^\alpha,\mathcal{C}^\beta} q_{\alpha \beta}} \, .
\end{equation}
The saddle-point equations trivially give $q_{\alpha \beta} = \langle \delta_{\mathcal{C}^\alpha,\mathcal{C}^\beta} \rangle_1$, where the average is computed with the single-site Hamiltonian ${\cal H}_1$ defined from the single-site partition function in Eq.~(\ref{eq:A_qab}) by $Z_1 [q_{\alpha \beta}] = {\rm Tr} \, e^{- {\cal H}_1}$.
Note that in the following we will make repeated use of the following identity:
\begin{equation} \label{eq:identity}
\delta_{\mathcal{C}^\alpha,\mathcal{C}^\beta} = \sum_{\mathcal{C}=1}^{2^M} \delta_{\mathcal{C}^\alpha,\mathcal{C}} \,\delta_{\mathcal{C}^\beta,\mathcal{C}} \, .
\end{equation}

\subsubsection{The Replica Symmetric solution}

We first consider a replica symmetric (RS) ansaz for the matrix $q_{\alpha \beta}$. Using Eq.~(\ref{eq:identity}) the single-site Hamiltonian ${\cal H}_1$ can be rewritten as
\begin{displaymath}
\mathcal{H}_1 =  n \, \frac{M \beta^2 q_0}{4} - \frac{M \beta^2 q_0}{4} \sum_{\mathcal{C}}  \sum_{\alpha , \beta} \delta_{\mathcal{C}^\alpha,\mathcal{C}} \, \delta_{\mathcal{C}^\beta,\mathcal{C}}  \, .
\end{displaymath}
We now introduce $2^M$ gaussian integrals to decouple the sum over replicas,
\begin{displaymath} 
e^{\frac{M \beta^2 q_0}{4} \sum_{\mathcal{C}}  \sum_{\alpha , \beta} 
\delta_{\mathcal{C}^\alpha,\mathcal{C}} \, \delta_{\mathcal{C}^\beta,\mathcal{C}}} = 
\int \prod_{\mathcal{C}=1}^{2^M} \bigg[ \frac{\textrm{d} 
z_{\mathcal{C}}}{\sqrt{2 \pi}} e^{-z_{\mathcal{C}}^2/2} \bigg] 
\exp \bigg( \sqrt{\frac{M \beta^2 q_0}{2} }\sum_{\mathcal{C}} 
\sum_\alpha \delta_{\mathcal{C}^\alpha,\mathcal{C}} \, z_{\mathcal{C}} \bigg) \, .
\end{displaymath}
The $n$ replicas are now totally decoupled and the trace over $\{ \mathcal{C}^\alpha \}$ is given by the product of $n$ independent traces over single replicas,
\begin{displaymath} 
Z_1 [q_0] = e^{-n \frac{M \beta^2 q_0}{4} } \int \prod_{\mathcal{C}=1}^{2^M} \bigg[ \frac{\textrm{d} z_{\mathcal{C}}}{\sqrt{2 \pi}} e^{-z_{\mathcal{C}}^2/2} \bigg]  \big [ \tilde Z (q_0,\{z_{\mathcal{C}} \}) \big ]^n \, ,
\qquad \textrm{where} \qquad
\tilde Z (q_0,\{z_{\mathcal{C}} \}) = \sum_{\mathcal{C}} 
 \exp \bigg( \sqrt{\frac{M \beta^2 q_0}{2} } z_{\mathcal{C}}  \bigg) \, .
 \end{displaymath} 
In the $n \to 0$ limit one has
\begin{displaymath}
- \ln Z_1 [q_0] = n \, \frac{M \beta^2 q_0}{4}  - n \int \prod_{\mathcal{C}=1}^{2^M} \bigg[ \frac{\textrm{d} z_{\mathcal{C}}}{\sqrt{2 \pi}} e^{-z_{\mathcal{C}}^2/2} \bigg]  \ln \tilde Z (q_0,\{z_{\mathcal{C}} \}) \, ,
\end{displaymath}
and the RS free energy per site reads  in the thermodynamic limit
\begin{displaymath}
f(q_0) = - \frac{M \beta^2}{8} - \frac{M \beta^2}{8} \, q_0^2 +  \frac{M \beta^2}{4} \, q_0
 -  \int \prod_{\mathcal{C}=1}^{2^M} \mathcal{D}z_{\mathcal{C}}
\ln  \tilde Z (q_0,\{z_{\mathcal{C}} \}) \, ,
\end{displaymath}
where $\mathcal{D}x = e^{-x^2/2} \textrm{d}x/\sqrt{2 \pi}$. By taking the derivative of the free energy with respect to $q_0$ we obtain the following saddle point equation:
\begin{equation}  \label{eq:SPRS}
q_0 = 1 - \sqrt{\frac{2}{M \beta^2 q_0}} \int \prod_{\mathcal{C}=1}^{2^M} \mathcal{D}z_{\mathcal{C}}
\left[ \sum_{\mathcal{C}} 
 z_{\mathcal{C}} \, e^{\sqrt{\frac{M \beta^2 q_0}{2} } z_{\mathcal{C}}} \bigg / 
 \sum_{\mathcal{C}} e^{\sqrt{\frac{M \beta^2 q_0}{2} } z_{\mathcal{C}}} \right] \, ,
 \end{equation}
which can be easily solved numerically.

\subsubsection{The 1-RSB solution}

In the following we introduce a 1-step replica-symmetry breaking (1-RSB) ansatz\cite{1RSB} for the matrix $q_{\alpha \beta}$ by considering $n/m$ blocks of $m$ replicas such that $q_{\alpha \beta} = q_1$ if $(\alpha, \beta)$ belong to the same block and $q_{\alpha \beta} = q_0$ if  $(\alpha, \beta)$ belong to different blocks. From Eq.~(\ref{eq:identity}) the single-site Hamiltonian ${\cal H}_1$ can be rewritten as
\begin{displaymath}
{\mathcal{H}}_1 =  n \, \frac{M \beta^2  q_1}{4} - \frac{M \beta^2 q_0}{4} \sum_{\mathcal{C}}  \sum_{\alpha , \beta} \delta_{\mathcal{C}^\alpha,\mathcal{C}} \, \delta_{\mathcal{C}^\beta,\mathcal{C}}  -  \frac{M \beta^2 (q_1 - q_0)}{4} \sum_{\mathcal{C}}  \sum_{\alpha , \beta}^\star \delta_{\mathcal{C}^\alpha,\mathcal{C}} \, \delta_{\mathcal{C}^\beta,\mathcal{C}} \, ,
\end{displaymath}
where $\sum^\star$ is the sum over all possible couples $(\alpha,\beta)$ belonging to the same block. We introduce $2^M$ gaussian integrals to decouple the first sum over replicas in the above expression. The $n/m$ blocks of replicas are now totally decoupled and the trace over $\{ \mathcal{C}^\alpha \}$ is given by the product of $n/m$ independent traces over the replica indices of each block. We thus find
\begin{displaymath}
Z_1 [q_1,q_0,m] = e^{-n \frac{M \beta^2 q_1}{4} } \int \prod_{\mathcal{C}=1}^{2^M} \mathcal{D} z_{\mathcal{C}}
\big [ Z_{\rm block} (q_1,q_0,m,\{z_{\mathcal{C}} \}) \big ]^{\frac{n}{m}} \, ,
\end{displaymath}
where
\begin{equation} \label{eq:Zblock}
Z_{\rm block} (q_1,q_0,m,\{z_{\mathcal{C}} \}) = \sum_{\{ \mathcal{C}^\alpha \}^\star} 
 \exp \bigg( \sqrt{\frac{M \beta^2 q_0}{2} }\sum_{\mathcal{C}} \sum_\alpha^\star \delta_{\mathcal{C}^\alpha,\mathcal{C}} \, z_{\mathcal{C}}  + \frac{M \beta^2 (q_1 - q_0)}{4} \sum_{\mathcal{C}} \sum_{\alpha , \beta}^\star \delta_{\mathcal{C}^\alpha,\mathcal{C}} \, \delta_{\mathcal{C}^\beta,\mathcal{C}} \bigg) \, .
 \end{equation} 
The trace over $\{ \mathcal{C}^\alpha \}^\star$ involves only one block of replicas and $\alpha=1,\ldots,m$. In the limit $n \to 0$,  $[ Z_{\rm block} (q_1,q_0,m,\{z_{\mathcal{C}} \}) ]^{n/m} \approx 1 + (n/m) \ln Z_{\rm block} (q_1,q_0,m,\{z_{\mathcal{C}} \})$, yielding
\begin{displaymath}
- \ln Z_1 [q_1,q_0,m] = n \, \frac{M \beta^2 q_1}{4}  - \frac{n}{m} \int \prod_{\mathcal{C}=1}^{2^M} \mathcal{D} z_{\mathcal{C}}  \ln Z_{\rm block} (q_1,q_0,m,\{z_{\mathcal{C}} \}) \, .
\end{displaymath}
The second sum in the right-hand side of Eq.~(\ref{eq:Zblock}) can again be decoupled by introducing $2^M$ additional gaussian integrals,
 \begin{displaymath}
Z_{\rm block} (q_1,q_0,m,\{z_{\mathcal{C}} \}) =  \int \prod_{\mathcal{C}=1}^{2^M} \mathcal{D} w_{\mathcal{C}} 
\left[\tilde Z (q_1,q_0,m,\{z_{\mathcal{C}} \},\{w_{\mathcal{C}} \})\right]^m \, ,
\end{displaymath}
where
\begin{displaymath}
\tilde Z (q_1,q_0,m,\{z_{\mathcal{C}} \},\{w_{\mathcal{C}} \}) = \sum_{\mathcal{C}} \exp 
\bigg( \sqrt{\frac{M \beta^2 q_0}{2} } z_{\mathcal{C}} + \sqrt{\frac{M \beta^2 (q_1-q_0)}{2} } w_{\mathcal{C}}  
\bigg) \, .
\end{displaymath}
Finally, we obtain the free energy per site in the thermodynamic limit as
\begin{displaymath}
\begin{split}
f(q_1,q_0,m) &= - \frac{M \beta^2}{8} + \frac{M \beta^2}{8} \left[ (m-1) q_1^2 - m q_0^2 \right] + \frac{M \beta^2}{4} \, q_1 \\
& \qquad 
- \frac{1}{m} \int \prod_{\mathcal{C}=1}^{2^M} \mathcal{D}z_{\mathcal{C}}
\ln \bigg \{ \int \prod_{\mathcal{C}=1}^{2^M} \mathcal{D}w_{\mathcal{C}}\left[ \tilde Z (q_1,q_0,m,\{z_{\mathcal{C}} \},\{w_{\mathcal{C}} \})\right]^m \bigg \} \, .
\end{split}
\end{displaymath}
The saddle-point equations are obtained by imposing that the derivatives of the free energy with respect to $q_1$, $q_0$, and $m$ vanish. Taking the derivative with respect to $q_1$ leads to
 \begin{displaymath}
 \begin{split}
& (m-1)q_1 + 1 - \sqrt{\frac{2}{M \beta^2 (q_1-q_0)}} \int \prod_{\mathcal{C}=1}^{2^M} 
\mathcal{D}z_{\mathcal{C}} \times \\
 & \qquad \qquad \Bigg \{
 \int \prod_{\mathcal{C}=1}^{2^M} \mathcal{D}w_{\mathcal{C}} 
 \Bigg[ \big[\tilde Z (q_1,q_0,m,\{z_{\mathcal{C}} \},\{w_{\mathcal{C}} \})\big]^{m-1}  
 \bigg[ \sum_{\mathcal{C}} w_{\mathcal{C}} \exp \Big( \sqrt{\frac{M \beta^2 q_0}{2} } 
 z_{\mathcal{C}} + \sqrt{\frac{M \beta^2 (q_1-q_0)}{2} } w_{\mathcal{C}}  \Big) \bigg] \Bigg ]\\
 &\qquad \qquad \qquad \qquad \bigg /
 \int \prod_{\mathcal{C}=1}^{2^M} \mathcal{D}w_{\mathcal{C}} 
 \big[\tilde Z (q_1,q_0,m,\{z_{\mathcal{C}} \},\{w_{\mathcal{C}} \})\big]^m  \Bigg \} = 0 \, .
 \end{split}
 \end{displaymath}
The derivative with respect to $q_0$ gives
 \begin{displaymath}
 \begin{split}
& - m q_0 - \sqrt{\frac{2}{M \beta^2 }} \int \prod_{\mathcal{C}=1}^{2^M} \mathcal{D}z_{\mathcal{C}} \times \\
 & \Bigg \{
 \int \prod_{\mathcal{C}=1}^{2^M} \mathcal{D}w_{\mathcal{C}} 
 \Bigg[ \big[\tilde Z (q_1,q_0,m,\{z_{\mathcal{C}} \},\{w_{\mathcal{C}} \})\big]^{m-1}  
 \bigg[ \sum_{\mathcal{C}} \Big(\frac{z_{\mathcal{C}}}{\sqrt{q_0}} -  
 \frac{w_{\mathcal{C}}}{\sqrt{q_1-q_0}} \Big) \exp \Big( \sqrt{\frac{M \beta^2  q_0}{2} } z_{\mathcal{C}} + 
 \sqrt{\frac{M \beta^2  (q_1-q_0)}{2} } w_{\mathcal{C}}  \Big) \bigg] \Bigg ]\\
 & \qquad \qquad 
 \bigg / \int \prod_{\mathcal{C}=1}^{2^M} \mathcal{D}w_{\mathcal{C}}  \big[\tilde Z (q_1,q_0,m,\{z_{\mathcal{C}} \},\{w_{\mathcal{C}} \})\big]^m \Bigg \} = 0 \, .
 \end{split}
 \end{displaymath}
Note that for $m=1$ this equation gives back Eq.~(\ref{eq:SPRS}). Finally, the derivative with respect to $m$ gives
\begin{equation} \label{eq:SPm}
\begin{split}
& \frac{M \beta^2 }{8} (q_1^2 - q_0^2)
+ \frac{1}{m^2} \int \prod_{\mathcal{C}=1}^{2^M} \mathcal{D}z_{\mathcal{C}}
\ln \bigg \{ \int \prod_{\mathcal{C}=1}^{2^M} \mathcal{D}w_{\mathcal{C}}  \big[\tilde Z (q_1,q_0,m,\{z_{\mathcal{C}} \},\{w_{\mathcal{C}} \})\big]^m \bigg \} \\
& \qquad \qquad
- \frac{1}{m}  \int \prod_{\mathcal{C}=1}^{2^M} \mathcal{D}z_{\mathcal{C}} 
\Bigg \{
 \int \prod_{\mathcal{C}=1}^{2^M} \mathcal{D}w_{\mathcal{C}} \big[ \big[\tilde Z (q_1,q_0,m,\{z_{\mathcal{C}} \},\{w_{\mathcal{C}} \})\big]^m 
 \ln  \tilde Z (q_1,q_0,m,\{z_{\mathcal{C}} \},\{w_{\mathcal{C}} \}) \big] 
 \\
& \qquad \qquad \qquad \qquad
 \bigg / \int \prod_{\mathcal{C}=1}^{2^M} \mathcal{D}w_{\mathcal{C}}  \big[\tilde Z (q_1,q_0,m,\{z_{\mathcal{C}} \},\{w_{\mathcal{C}} \})\big]^m \Bigg \} = 0 \, .
 \end{split}
\end{equation}
In order to find the thermodynamic glass transition, one thus needs to solve numerically Eq.~(\ref{eq:SPRS}), which yields the value of $q_0$, and Eq.~(\ref{eq:SPm}) for $m=1$, which gives the value of $q_1$ such that $m=1$ is an extremum of the free-energy, and finally check whether $q_1 \neq 
q_0$. After some simple algebra Eq.~(\ref{eq:SPm}) for $m=1$ can be rewritten in a simpler form:
\begin{displaymath} 
\begin{split}
& q_1^2 - q_0^2
+ 2 (q_1 - q_0) + \frac{8}{M \beta^2}  \int \prod_{\mathcal{C}=1}^{2^M} \mathcal{D}z_{\mathcal{C}}
\ln \bigg( \sum_{\mathcal{C}} e^{\sqrt{\frac{M \beta^2  q_0}{2} } z_{\mathcal{C}}} \bigg) \\
& \qquad 
- \, \frac{8 e^{-M \beta^2  (q_1-q_0)/4}}{M \beta^2}
 \int \prod_{\mathcal{C}=1}^{2^M} \mathcal{D}z_{\mathcal{C}} 
\, \frac
 {\int \prod_{\mathcal{C}=1}^{2^M} \mathcal{D}w_{\mathcal{C}} 
 \big[\tilde Z (q_1,q_0,m,\{z_{\mathcal{C}} \},\{w_{\mathcal{C}} \}) 
 \ln  \tilde Z (q_1,q_0,m,\{z_{\mathcal{C}} \},\{w_{\mathcal{C}} \}) \big] }
{\sum_{\mathcal{C}} e^{\sqrt{\frac{M \beta^2 q_0}{2} } z_{\mathcal{C}}}} = 0 \, .
\end{split}
\end{displaymath}
The numerical solutions of the 1-RSB equations for $M=3$ corresponds to the black curve (circles) in Fig.~\ref{fig:overlap}, showing a transition (RFOT) for $\beta_K \approx 2.5$.

\subsection{Construction of the effective theory} \label{app:REMFC-Heff}

In the following we apply the procedure described in Sec.~\ref{sec:stage} to construct the effective theory of the model. To this aim we consider $n+1$ replicas of the system and compute the replicated action for a fixed overlap field $\{ p_a^i \}$ of the replicas $a=1,\ldots,n$ with a given reference configuration $\{ {\cal C}_i^0 \}$. Note that $p_a^i = 1$ only if $\mathcal{C}_i^a = \mathcal{C}_i^0$  and is zero otherwise. As already mentioned, we will consider the temperature range $T_d \le T \le T_K$, where we can use the annealed approximation to average over the random energies $E_{ i j }$. The replicated action is given in Eq.~(\ref{eq:Srep-REMFC}) of the main text, where the Kronecker $\delta$'s in the exponential can be rewritten in terms of the overlap variables as in Eq.~(\ref{eq:sum}). As discussed in the main text, if $\mathcal{C}_i^a = \mathcal{C}_i^0$ and $\mathcal{C}_i^b = \mathcal{C}_i^0$ ({\it i.e.}, $p_a^i = p_b^i = 1$), then $\mathcal{C}_i^b = \mathcal{C}_i^a$.  Similarly, if $\mathcal{C}_i^a = \mathcal{C}_i^0$ and $\mathcal{C}_i^b \neq \mathcal{C}_i^0$ ({\it i.e.}, $p_a^i = 1$ and $p_b^i = 0$) then $\mathcal{C}_i^b \neq \mathcal{C}_i^a$. The same is true, of course, if $\mathcal{C}_i^a \neq \mathcal{C}_i^0$ and $\mathcal{C}_i^b = \mathcal{C}_i^0$. The only undetermined case corresponds to $\mathcal{C}_i^a \neq \mathcal{C}_i^0$ and $\mathcal{C}_i^b \neq \mathcal{C}_i^0$.

\subsubsection{The average effective action: First cumulant} \label{app:REMFC_Eff_1}

We first focus on the first cumulant (1-replica action) $\mathcal S_1 [\{p^i\}]$. It is then sufficient to set all replica fields equal, $p_a^i =p^i$ $\forall \, a=1, \cdots,n$ and $\forall \, i$, keep only the term of order $n$ in the expression of $\mathcal S_{\rm rep}[\{p_a^i\}]$, and take the limit $n \to 0$ in the end, as in the standard replica trick. In order to do this we set the overlap profile with the reference configuration for all replicas to be $1$ on the first $cN$ sites (\ie $p_a^i = 1$ for $i=1,\ldots,cN$, $\forall a$) and $0$ on all the other $(1-c)N$ sites (\ie $p_a^i = 0$ for $i=cN+1,\ldots,N$, $\forall a$).

For the chosen overlap profile Eq.~(\ref{eq:sum}) becomes
\[
\beal
\sum_{i \neq j} \Big[ 1 + n + 2 n p^i p^j + \sum_{a\neq b} \delta_{\mathcal{C}_i^a,\mathcal{C}_i^b} \delta_{\mathcal{C}_j^a,\mathcal{C}_j^b}
\Big] = & (1 + n) N (N-1) + 2 n c N (cN - 1) + n (n-1) c N (c N - 1) \\
& \qquad \qquad  + 2 c N \sum_{a \neq b} \sum_i^\star 
\delta_{\mathcal{C}_i^a,\mathcal{C}_i^b} + \sum_{a \neq b} \Big[ 
\Big( \sum_i^\star \delta_{\mathcal{C}_i^a,\mathcal{C}_i^b} \Big)^2 - \sum_i^\star \delta_{\mathcal{C}_i^a,\mathcal{C}_i^b} \Big] \\
\approx & (1 + n) N^2 + n (n + 1) c^2 N^2 + 2 c N \sum_{a \neq b} \sum_i^\star \delta_{\mathcal{C}_i^a,\mathcal{C}_i^b} 
+ \sum_{a \neq b} \Big( \sum_i^\star \delta_{\mathcal{C}_i^a,\mathcal{C}_i^b} \Big)^2 \, ,
\eal
\]
where in going to the last line we have thrown away all the sub-extensive diagonal ($i=j$) terms. The sum $\sum_i^\star$ represents the sum over the sites $i = cN + 1, \ldots, N$ where $p^i=0$. Inserting this expression into Eq.~(\ref{eq:Srep-REMFC}) yields
\[
e^{-n {\cal S}_1 (c)} = \frac{e^{\frac{N \beta^2 M}{8} \left [ 1 + n + n (n+1) c^2 \right ]}}{\overline{Z}} 
\sum_{\{ \mathcal{C}_i^\alpha \}} e^{\frac{N \beta^2 M}{8} \left[ 2 c \sum_{a \neq b} \frac{1}{N} 
\sum_i^\star \delta_{\mathcal{C}_i^a,\mathcal{C}_i^b} + \sum_{a \neq b} \left( \frac{1}{N} 
\sum_i^\star \delta_{\mathcal{C}_i^a,\mathcal{C}_i^b} \right)^2 \right]}
\prod_{a,i} 
\delta_{p_a^i,\delta_{\mathcal{C}_i^0,\mathcal{C}_i^a}}
\,. 
\]
The sum over the reference configuration ${\cal C}_i^0$ simply gives $2^{NM}$. On the first $cN$ sites we have that $p^i=1$ and then ${\cal C}_i^a = {\cal C}_i^0$ for all $a$. We thus obtain
\[
e^{-n {\cal S}_1 (c)} = \frac{e^{N M \left \{ \ln 2 + \frac{\beta^2}{8} \left [ 1 + n + n (n+1) c^2 \right ] \right \} }}{\overline{Z}}
\sum_{\{ \mathcal{C}_i^a \}_\star } e^{\frac{\beta^2 M}{4} \sum_{a < b} \left [ 2 c \sum_i^\star \delta_{\mathcal{C}_i^a,\mathcal{C}_i^b}
+ \frac{1}{N} \left( \sum_i^\star \delta_{\mathcal{C}_i^a,\mathcal{C}_i^b} \right)^2 \right] } \, ,
\]
where the trace $\sum_{\{ \mathcal{C}_i^a \}_\star }$ represents the sum over all the $2^M - 1$ configurations $\mathcal{C}_i^a$ that are different from the reference one on the $(c-1)N$ sites where $p^i=0$. One can now introduce the overlaps $q_{ab}$ by performing the usual Hubbard-Stratonovich transformations,
\[
e^{\frac{\beta^2 M}{4 N} \sum_{a < b} \left( \sum_i^\star \delta_{\mathcal{C}_i^a,\mathcal{C}_i^b} \right)^2}
= \Big( \frac{N \beta^2 M}{4 \pi} \Big)^{n(n-1)/4} \int \prod_{a < b} 
{\rm d} q_{ab} \, e^{-\frac{N \beta^2 M}{4} \sum_{a < b} q_{ab}^2 + \frac{\beta^2 M}{2} \sum_{a < b} q_{ab} 
\sum_i^\star \delta_{\mathcal{C}_i^a,\mathcal{C}_i^b}} \, .
\]
At this point one can easily compute the trace over the configurations $\sum_{\{ \mathcal{C}_i^a \}_\star }$, thanks to the fact that the sites are decoupled (the annealed partition function of the model in the denominator only yields an unimportant constant term): 
\[
\beal
e^{-n {\cal S}_1 (c)} & = e^{\frac{n N \beta^2 M}{8} \left [  1 + (n+1) c^2 \right ] } \Big( \frac{N \beta^2 M}{4 \pi} \Big)^{n(n-1)/4}
\int \prod_{a < b} {\rm d} q_{ab} \, e^{- N A [q_{ab}]} \, ,\\
A [q_{ab}] & = \frac{M \beta^2}{4} \sum_{a < b} q_{ab}^2 - (1 - c) \ln Z_1 \, , \\
Z_1 & = \sum_{\{ {\cal C}^a \}_\star} e^{\frac{\beta^2 M}{2} \sum_{a < b} (q_{ab} + c) 
\delta_{\mathcal{C}^a,\mathcal{C}^b}} \, .
\eal
\]
In the thermodynamic limit ($N \to \infty$) the integral in the above expression can be performed via a saddle-point method, which gives $q_{ab} = (1 - c) \langle \delta_{\mathcal{C}^a,\mathcal{C}^b} \rangle_1$, the average $\langle \cdots \rangle_1$ being performed with the single-site Hamiltonian $- {\cal H}_1 = \sum_{a < b} (q_{ab} + c) \delta_{\mathcal{C}^a,\mathcal{C}^b}$.

We now introduce a RS ansatz for the overlap matrix, $q_{ab} = q_0$. Using the identity~(\ref{eq:identity}), $\delta_{\mathcal{C}^a,\mathcal{C}^b} = \sum_{\mathcal{C}}^\star \delta_{\mathcal{C}^a,\mathcal{C}} \, \delta_{\mathcal{C}^b,\mathcal{C}}$, the single-site Hamiltonian can be rewritten as
\[
- {\cal H}_1 = - \frac{n \beta^2 M}{4} (q_0 + c) + \frac{\beta^2 M}{4} (q_0 + c) \sum_{\mathcal{C}}^\star \Big(
\sum_a \delta_{\mathcal{C}^a,\mathcal{C}} \Big)^2 \, ,
\]
The replicas can again be decoupled via Hubbard-Stratonovich transformations, yielding
\[
\ln Z_1 = - \frac{n \beta^2 M}{4} (q_0 + c) + \ln 
\int \prod_{{\cal C}}^\star \frac{{\rm d} z_{{\cal C}}}{\sqrt{2 \pi}} e^{- \frac{z_{{\cal C}}^2}{2}}
\Bigg[ \sum_{{\cal C}}^\star e^{\sqrt{\frac{\beta^2 M (q_0 + c)}{2}} z_{{\cal C}}} \Bigg]^n \, .
\]
In the $n \to 0$ limit one thus has
\[
A[q_0] = \frac{n(n-1) \beta^2 M q_0^2}{8} + \frac{n (1 - c) \beta^2 M (q_0 + c)}{4} - n (1 - c) 
\, \overline{\ln \Bigg[ \sum_{{\cal C}}^\star e^{\sqrt{\frac{\beta^2 M (q_0 + c)}{2}} z_{{\cal C}}} \Bigg]}
\, ,
\]
where the average $\overline{[ f(\vec{z}_{\cal C})]}$ is defined over the gaussian measure $\overline{[ f(\vec{z}_{\cal C})]} \equiv \int \prod_{{\cal C}}^\star \left[ \frac{{\rm d}z_{\cal C}}{\sqrt{2 \pi}} \, e^{- z_{\cal C}^2/2} \right] f(\vec{z}_{\cal C})$. Putting all the results together in the $n \to 0$ limit (and neglecting subleading terms in the $N \to \infty$ limit), one gets Eq.~(\ref{eq:S1c}) given in the main text, where the overlap $q_0$ must satisfy the self-consistent equation~(\ref{eq:q0}). The solution of such a saddle-point equation can be developed in powers of $c$ as $q_0 \approx q_{0,0} +  c q_{0,1} + c^2 q_{0,2} + \ldots$, 
which, when inserted back into Eqs.~(\ref{eq:q0}) and~(\ref{eq:S1c}), allows one to obtain the exact expansion of ${\cal S}_1 (c)$ in powers of $c$. In order to provide analytic expressions of the coupling constants $K_n$ appearing in Eq.~(\ref{eq:Heff_expansion}), we perform the expansion of ${\cal S}_1 (c)$ in powers
of $c$ when $M \gg 1$. After expanding the exponentials of Eqs.~(\ref{eq:S1c}) and (\ref{eq:q0}) up to the eighth order in $\sqrt{M \beta^2 (q_0 + c)/2}$, we obtain
\begin{equation} \label{eq:q0-Kac}
q_0 \approx \frac{1}{2^M} + \frac{M \beta^2 -2}{2^{M+1}} c + \frac{M \beta^2 (M \beta^2 - 4)}{2^{M+3}} c^2 + \frac{M^2 \beta^4(M \beta^2 - 6)}{3 \cdot 2^{M+4}} c^3 - \frac{M^3 \beta^6}{3 \cdot 2^{M+4}} 
c^4 + \ldots \, .
\end{equation}
Inserting this expression into Eq.~(\ref{eq:S1c}) and re-expressing the powers of $c$ (up to fourth order) as effective one-, two-, three-, and four-body interactions, we find an effective Hamiltonian of the form given in Eq.~(\ref{eq:Heff_expansion}) with parameters given in Eq.~(\ref{eq:Kn}).

\subsubsection{Fluctuations of the effective action: Second cumulant} 
\label{app:REMFC_Eff_2}

We now turn to the computation of the second cumulant (2-replica action) $\mathcal S_2[\{p_1^i,p_2^i\}]$. To do this we divide the $n$ constrained replicas into two groups of $n_1$ and $n_2$ replicas respectively. The most generic overlap profile can be obtained by dividing the sites in four groups, denoted $(1)$, $(2)$, $(12)$, and $(0)$, such that on the $c_1 N$ sites belonging to the group $(1)$ $p_1^i=1$ and $p_2^i=0$, on the $c_2 N$ sites belonging to the group $(2)$ $p_1^i=0$ and $p_2^i=1$, on the $c_{12} N$ sites belonging to the group $(12)$ $p_1^i = p_2^i = 1$, and on the $c_0 N$ sites belonging to the group $(0)$ $p_1^i = p_2^i = 0$ (with $c_0 = 1-c_1-c_2-c_{12}$). The second cumulant can be computed by keeping only the terms of order $n_1 n_2$ in the expression of the replicated action, and taking the limit $n_1 , n_2 \to 0$ [see Eq.~(\ref{eq:FRS})]. The first terms of the right-hand side of Eq.~(\ref{eq:sum}) thus become (neglecting subleading corrections)
\[
\sum_{i \neq j} \Big[ 1 + n_1 + n_2 + 2 \sum_a p_a^i p_a^j \Big] \approx N^2 \big[ 1 + n_1 + n_2 + 2 n_1 (c_1 + c_{12})^2
+ 2 n_2 (c_2 + c_{12})^2 \Big] \, .
\]
On the other hand, the second term of the right-hand side of Eq.~(\ref{eq:sum}) reads (neglecting again subleading corrections)
\[
\beal
 \sum_{i \neq j} \sum_{a\neq b} \delta_{\mathcal{C}_i^a,\mathcal{C}_i^b} \delta_{\mathcal{C}_j^a,\mathcal{C}_j^b}
& \approx N^2 \Big[ (c_1 + c_{12})^2 n_1 (n_1 - 1) + (c_2 + c_{12})^2 n_2 (n_2 - 1) + 2 c_{12}^2 n_1 n_2 \Big] \\
& + 2 (c_1 + c_{12}) N \sumabu \sumidz \dCaCbi + 2 (c_2 + c_{12}) N \sumabd \sumiuz \dCaCbi 
+ 4 c_{12} N \sumau \sumbd \sumiz \dCaCbi \\
&+ \sumabu \Bigg( \sumidz \dCaCbi \Bigg)^2  
+ \sumabd \Bigg( \sumiuz \dCaCbi \Bigg)^2  
+ 2 \sumau \sumbd \Bigg( \sumiz \dCaCbi \Bigg)^2 \, ,
\eal
\]
where $\sum_a^{(1),(2)}$ denotes the sum over the constrained replicas belonging, respectively, to the first group ($a = 1, \ldots, n_1$) or to the second group ($a = 1 + n_1, \ldots, n_1 + n_2$), while $\sum_i^{(s)}$ denotes the sum over $i$ belonging to the $s$-th group of sites (with $s=1$, $2$, $0$, or $12$). We can now perform the sum over the reference configuration ${\cal C}_i^0$, which simply gives $2^{MN}$. On the $c_1 N$ sites belonging to the group $(1)$, $p_a^i=1$ for all $a$ in the first group of replicas and $p_a^i=0$ for all $a$ in the second group. Hence ${\cal C}_i^a = {\cal C}_i^0$ for all $a=1, \ldots, n_1$, whereas ${\cal C}_i^a \neq {\cal C}_i^0$ for all $a=1+n_1, \ldots, n_1+n_2$. Similarly, on the $c_2 N$ sites belonging to the group $(2)$, $p_a^i=0$ for all $a$ in the first group of replicas and $p_a^i=1$ for all $a$ in the second group. Hence ${\cal C}_i^a \neq {\cal C}_i^0$ for all $a=1, \ldots, n_1$, whereas ${\cal C}_i^a = {\cal C}_i^0$ for all $a=1+n_1, \ldots, n_1+n_2$. On the $c_{12} N$ sites belonging to the group $(12)$, $p_a^i=1$ for all replicas. Thus ${\cal C}_i^a = {\cal C}_i^0$ for all $a=1, \ldots, n_1+n_2$. Finally, on the $c_{0} N$ sites belonging to the group $(0)$, $p_a^i=0$ for all replicas. Thus ${\cal C}_i^a \neq {\cal C}_i^0$ for all $a=1, \ldots, n_1+n_2$. In consequence, the trace over the configuration $\sum_{\{ \mathcal{C}_i^a \}}$ consists in summing over all possible $2^M - 1$ configurations different from the reference one for $a$ belonging to the second group of replicas on the sites of group $(1)$, to the first group of replicas on the sites of group $(2)$, and on all replicas on the sites of group $(0)$. In the following we will denote this sum as $\sum_{\{ \mathcal{C}_i^a \}_\star}$.

At this point we introduce several overlaps, $q_{ab}^{[1]}$, $q_{ab}^{[2]}$, $q_{ab}^{[12]}$, via the usual Hubbard-Stratonovich transformations that allow one to decouple different sites,
\[
\beal
e^{\frac{\beta^2 M}{4 N} \sum_{a < b}^{(1)} \left( \sumidz \dCaCbi \right)^2}
&= \Big( \frac{N \beta^2 M}{4 \pi} \Big)^{n_1(n_1-1)/4} \int \prod_{a < b}^{(1)} 
{\rm d} q_{ab}^{[1]} \, e^{-\frac{N \beta^2 M}{4} \sum_{a < b}^{(1)} \left (q_{ab}^{[1]} \right)^2 + 
\frac{\beta^2 M}{2} \sum_{a < b}^{(1)} q_{ab}^{[1]} 
\sumidz \dCaCbi} \, , \\
e^{\frac{\beta^2 M}{4 N} \sum_{a < b}^{(2)} \left( \sumiuz \dCaCbi \right)^2}
&= \Big( \frac{N \beta^2 M}{4 \pi} \Big)^{n_2(n_2-1)/4} \int \prod_{a < b}^{(2)} 
{\rm d} q_{ab}^{[2]} \, e^{-\frac{N \beta^2 M}{4} \sum_{a < b}^{(2)} \left (q_{ab}^{[2]} \right )^2 + \frac{\beta^2 M}{2} 
\sum_{a < b}^{(2)} q_{ab}^{[2]} 
\sumiuz \dCaCbi }\, , \\
e^{\frac{\beta^2 M}{4 N} \sumau \sumbd \left( \sumiz \dCaCbi \right)^2}
&= \Big( \frac{N \beta^2 M}{4 \pi} \Big)^{n_1 n_2/2} \int \!\! \prod_{a \in (1);b \in (2) } \!\!
{\rm d} q_{ab}^{[12]} \, e^{-\frac{N \beta^2 M}{4} \sumau \sumbd \left (q_{ab}^{[12]} \right )^2 + \frac{\beta^2 M}{2} 
\sumau \sumbd q_{ab}^{[12]} 
\sumiz \dCaCbi }\, .
\eal
\]
Neglecting all the subleading and irrelevant terms, we can now rewrite the replicated action in the following way:
\begin{equation} \label{eq:A2}
\beal
e^{- {\cal S}_{\rm rep} [c_1,c_2,c_{12}]} & 
= e^{\frac{N \beta^2 M}{8} \left [ n_1 + n_2 + n_1 (n_1+1) (c_1 + c_{12})^2  + n_2 (n_2+1) (c_2 + c_{12})^2
+ 2 c_{12}^2 n_1 n_2 \right ] } \\
& \qquad \qquad \qquad \qquad \times 
\int \prod_{a < b}^{(1)} {\rm d} q_{ab}^{[1]} \prod_{a < b}^{(2)} {\rm d} q_{ab}^{[2]} \prod_{a \in (1);b \in (2) } \!\!
{\rm d} q_{ab}^{[12]}  \, e^{- N A [q_{ab}^{[1]},q_{ab}^{[2]},q_{ab}^{[12]}]} \, ,\\
A [q_{ab}^{[1]},q_{ab}^{[2]},q_{ab}^{[12]}] & = \frac{M \beta^2}{4} 
\Big[ \sum_{a < b}^{(1)} \big(q_{ab}^{[1]}\big)^2 + \sum_{a < b}^{(2)} \big(q_{ab}^{[2]}\big)^2  
+ \sumau \sumbd \big(q_{ab}^{[12]}\big)^2 \Big]
- c_1 \ln Z_1 - c_2 \ln Z_2 - (1 - c_1 - c_2 - c_{12}) \ln Z_0 \, , \\
Z_1 & = \sum_{\{ {\cal C}^a \}_\star} e^{\frac{\beta^2 M}{2} \sum_{a < b}^{(2)} (q_{ab}^{[2]} + c_2 + c_{12}) 
\delta_{\mathcal{C}^a,\mathcal{C}^b}} \, , \\
Z_2 & = \sum_{\{ {\cal C}^a \}_\star} e^{\frac{\beta^2 M}{2} \sum_{a < b}^{(1)} (q_{ab}^{[1]} + c_1 + c_{12}) 
\delta_{\mathcal{C}^a,\mathcal{C}^b}} \, , \\
Z_0 & = \sum_{\{ {\cal C}^a \}_\star} e^{\frac{\beta^2 M}{2} \left[ 
\sum_{a < b}^{(1)} (q_{ab}^{[1]} + c_1 + c_{12}) \delta_{\mathcal{C}^a,\mathcal{C}^b} 
+ \sum_{a < b}^{(2)} (q_{ab}^{[2]} + c_2 + c_{12}) \delta_{\mathcal{C}^a,\mathcal{C}^b}
+ \sumau \sumbd (q_{ab}^{[12]} + c_{12}) \delta_{\mathcal{C}^a,\mathcal{C}^b} \right]} \, .
\eal
\end{equation}
The integrals over the overlaps can be performed at the saddle point in the limit $N \to \infty$, which gives
\[
\beal
q_{a_1 b_1}^{[1]} &= c_2 \langle \delta_{\mathcal{C}^{a_1},\mathcal{C}^{b_1}} \rangle_2 + (1 - c_1 - c_2 - c_{12}) 
\langle \delta_{\mathcal{C}^{a_1},\mathcal{C}^{b_1}} \rangle_0 \, \\
q_{a_2 b_2}^{[2]} &= c_1 \langle \delta_{\mathcal{C}^{a_2},\mathcal{C}^{b_2}} \rangle_1 + (1 - c_1 - c_2 - c_{12}) 
\langle \delta_{\mathcal{C}^{a_2},\mathcal{C}^{b_2}} \rangle_0 \, \\
q_{a_1 b_2}^{[12]} &= (1 - c_1 - c_2 - c_{12}) 
\langle \delta_{\mathcal{C}^{a_1},\mathcal{C}^{b_2}} \rangle_0 \, ,
\eal
\]
where the indices $a_1$ and $b_1$ (resp., $a_2$ and $b_2$) belong to the first (resp., second) group of replicas ({\it i.e.}, $a_1,b_1 = 1, \ldots , n_1$ and $a_2,b_2 = 1+n_1, \ldots , n_1+n_2$), and the averages are performed over the single-site Hamiltonians ${\cal H}_1$, ${\cal H}_2$, and ${\cal H}_0$, corresponding to (minus) the arguments of the exponentials appearing in the expressions of $Z_0$, $Z_1$, and $Z_2$ in Eq.~(\ref{eq:A2}). We now introduce a RS ansatz for the overlaps, $q_{ab}^{[1]} = q_1$, $q_{ab}^{[2]} = q_2$, $q_{ab}^{[12]} = q_{12}$, $\forall a,b$, which is justified for $T\geq T_K$, at least for small $c_1$, $c_2$ and $c_{12}$. Using once more the identity in Eq.~(\ref{eq:identity}), $\delta_{\mathcal{C}^a,\mathcal{C}^b} = \sum_{\mathcal{C}}^\star \delta_{\mathcal{C}^a,\mathcal{C}} \, \delta_{\mathcal{C}^b,\mathcal{C}}$, the single-site Hamiltonians can be rewritten as
\[
\beal
- {\cal H}_1 & = - \frac{n_2 \beta^2 M}{4} (q_2 + c_2 + c_{12}) + \frac{\beta^2 M}{4} (q_2 + c_2 + c_{12}) \sum_{\mathcal{C}}^\star \Big(
\sumad \delta_{\mathcal{C}^a,\mathcal{C}} \Big)^2 \, , \\
- {\cal H}_2 & = - \frac{n_1 \beta^2 M}{4} (q_1 + c_1 + c_{12}) + \frac{\beta^2 M}{4} (q_1 + c_1 + c_{12}) \sum_{\mathcal{C}}^\star \Big(
\sumau \delta_{\mathcal{C}^a,\mathcal{C}} \Big)^2 \, , \\
- {\cal H}_0 & = - {\cal H}_1 - {\cal H}_2 + \frac{\beta^2 M}{2} (q_{12} + c_{12}) \sum_{\mathcal{C}}^\star
\sumau \delta_{\mathcal{C}^a,\mathcal{C}} \sumbd \delta_{\mathcal{C}^b,\mathcal{C}} \, .
\eal
\]
The replicas can again be decoupled via Hubbard-Stratonovich transformations, yielding (in the limit $n_1,n_2 \to 0$ and keeping only terms up to second order in the number of replicas)
\begin{equation} \label{eq:Z1Z2}
\beal
\ln Z_1 &= - \frac{n_2 \beta^2 M}{4} (q_2 + c_2 + c_{12}) + 
n_2 \, \overline{ \Bigg[ \ln \Bigg( \sum_{{\cal C}}^\star e^{\sqrt{\frac{\beta^2 M (q_2 + c_2 + c_{12})}{2}} z_{{\cal C}}} \Bigg) \Bigg]} \\
& \qquad \qquad + \frac{n_2^2}{2} \left \{ 
\overline{\Bigg[ \ln \Bigg( \sum_{{\cal C}}^\star e^{\sqrt{\frac{\beta^2 M (q_2 + c_2 + c_{12})}{2}} z_{{\cal C}}} \Bigg) \Bigg]^2} - 
\left[ \overline{\Bigg[ \ln \Bigg( \sum_{{\cal C}}^\star e^{\sqrt{\frac{\beta^2 M (q_2 + c_2 + c_{12})}{2}} z_{{\cal C}}} \Bigg) \Bigg]}  \right ]^2 \right \}\, , \\
\ln Z_2 &= - \frac{n_1 \beta^2 M}{4} (q_1 + c_1 + c_{12}) + n_1 \, \overline{\Bigg[ \ln 
\Bigg( \sum_{{\cal C}}^\star e^{\sqrt{\frac{\beta^2 M (q_1 + c_1 + c_{12})}{2}} z_{{\cal C}}} \Bigg) \Bigg]} \\
& \qquad \qquad + \frac{n_1^2}{2} \left \{
\overline{\Bigg[ \ln \Bigg( \sum_{{\cal C}}^\star e^{\sqrt{\frac{\beta^2 M (q_1 + c_1 + c_{12})}{2}} z_{{\cal C}}} \Bigg) \Bigg]^2}
- \left[ 
\overline{\Bigg[ \ln \Bigg( \sum_{{\cal C}}^\star e^{\sqrt{\frac{\beta^2 M (q_1 + c_1 + c_{12})}{2}} z_{{\cal C}}} \Bigg) \Bigg]} \right]^2 \right \} \, , \\
\eal
\end{equation}
where, as before, the averages $\overline{[ f(\vec{z}_{\cal C})]}$ are defined over the gaussian measure $\overline{[ f(\vec{z}_{\cal C})]} \equiv \int \prod_{{\cal C}}^\star \left[ \frac{{\rm d}z_{\cal C}}{\sqrt{2 \pi}} \, e^{- z_{\cal C}^2/2} \right] f(\vec{z}_{\cal C})$.

The computation of $Z_0$ is slightly more involved. After introducing $2 (2^M - 1)$ $\delta$-functions enforcing $x_{\cal C} = i \sqrt{\frac{\beta^2 M (q_1 + c_1 + c_{12})}{2}} \sumau \delta_{\mathcal{C}^a,\mathcal{C}}$ and $y_{\cal C} = i \sqrt{\frac{\beta^2 M (q_2 + c_2 + c_{12})}{2}} \sumad \delta_{\mathcal{C}^a \mathcal{C}}$ in the RS ansatz, $Z_0$ becomes
\[
\beal
Z_0 & = e^{-\frac{\beta^2 M}{4} \left[ n_2 (q_2 + c_2 + c_{12}) + n_1 (q_1 + c_1 + c_{12}) \right]} \\
& \times \sum_{\{ {\cal C}^a \}_\star} \prod_{\cal C}^\star \Bigg \{ 
\int_{- \infty}^{\infty} {\rm d} x_{\cal C} \, \delta \Bigg( i \sqrt{\frac{\beta^2 M (q_1 + c_1 + c_{12})}{2}}
\sumau \delta_{\mathcal{C}^a,\mathcal{C}} - x_{\cal C} \Bigg)
\int_{- \infty}^{\infty} {\rm d} y_{\cal C} \, \delta \Bigg( i \sqrt{\frac{\beta^2 M (q_2 + c_2 + c_{12})}{2}}  
\sumad \delta_{\mathcal{C}^a,\mathcal{C}} - y_{\cal C} \Bigg) \\ 
& \qquad \qquad \qquad \times  \exp \left ( 
- \frac{x_{\cal C}^2}{2} - 
\frac{y_{\cal C}^2}{2} - \frac{q_{12} + c_{12}}{\sqrt{(q_1 + c_1 + c_{12}) (q_2 + c_2 + c_{12})}}
x_{\cal C} y_{\cal C} 
\right ) \Bigg \} \, .
\eal
\]
Using the integral representation of the $\delta$-function, $\delta(x-x_0) = \int_{- \infty}^{+ \infty} {\rm d} \hat x \, e^{- i \hat x(x - x_0)}$, integrating over $x_{\cal C}$ and $y_{\cal C}$, and performing the trace over configurations, one then easily finds
\[
\beal
Z_0 & = e^{-\frac{\beta^2 M}{4} \left[ n_2 (q_2 + c_2 + c_{12}) + n_1 (q_1 + c_1 + c_{12}) \right]}  
\int \prod_{\cal C}^\star \left[ {\rm d} \hat x_{\cal C} \, {\rm d} \hat y_{\cal C} \frac{2 \pi}{\sqrt{1 - \gamma^2}}
\, e^{- \frac{1}{1 - \gamma^2} \left( \frac{\hat x_{\cal C}^2}{2} + \frac{\hat y_{\cal C}^2}{2} - 
\gamma \hat x_{\cal C} \hat y_{\cal C} \right)} \right] \\
& \qquad \qquad \qquad \times 
\Bigg( \sum_{{\cal C}}^\star e^{ \sqrt{\frac{\beta^2 M (q_1 + c_1 + c_{12})}{2}} \hat x_{\cal C}} \Bigg)^{n_1} 
\Bigg( \sum_{{\cal C}}^\star e^{ \sqrt{\frac{\beta^2 M (q_2 + c_2 + c_{12})}{2}} \hat y_{\cal C}} \Bigg)^{n_2} \, ,
\eal
\]
where
\[
\gamma = \frac{q_{12} + c_{12}}{\sqrt{(q_1 + c_1 + c_{12}) (q_2 + c_2 + c_{12})}} \, .
\]
Note that $\gamma$ must be less then one ({\it i.e.}, $q_{12} < q_1 + c_1$ and $q_{12} < q_2 + c_2$) for the Gaussian integrals to be well defined. We will find at the end of the computation that this is indeed the case. Expanding the logarithm of $Z_0$ in powers of $n_1$ and $n_2$ and keeping only terms up to second order we obtain
\begin{equation} \label{eq:Z0}
\beal
\ln Z_0 &= - \frac{\beta^2 M}{4} \left[ n_2 (q_2 + c_2 + c_{12}) + n_1 (q_1 + c_1 + c_{12}) \right] + \ln ( 4 \pi^2 ) \\
& + n_1 \, \overline{\Bigg[\ln \Bigg( \sum_{{\cal C}}^\star e^{ \sqrt{\frac{\beta^2 M (q_1 + c_1 + c_{12})}{2}} \hat x_{\cal C}} \Bigg) \Bigg]}
+ n_2 \, \overline{\Bigg[ \ln \Bigg( \sum_{{\cal C}}^\star e^{ \sqrt{\frac{\beta^2 M (q_2 + c_2 + c_{12})}{2}} \hat y_{\cal C}} \Bigg) \Bigg]} \\
& + \frac{n_1^2}{2} \left \{ \overline{\Bigg[ \ln \Bigg( \sum_{{\cal C}}^\star e^{ \sqrt{\frac{\beta^2 M (q_1 + c_1 + c_{12})}{2}} \hat x_{\cal C}} \Bigg) \Bigg]^2}
- \left [ \overline{\Bigg[ \ln \Bigg( \sum_{{\cal C}}^\star e^{ \sqrt{\frac{\beta^2 M (q_1 + c_1 + c_{12})}{2}} \hat x_{\cal C}} \Bigg) \Bigg]} \right]^2 \right \} \\
& + \frac{n_2^2}{2} \left \{ \overline{\Bigg[ \ln \Bigg( \sum_{{\cal C}}^\star e^{ \sqrt{\frac{\beta^2 M (q_2 + c_2 + c_{12})}{2}} \hat y_{\cal C}} \Bigg) \Bigg]^2}
- \left[ \overline{\Bigg[ \ln \Bigg( \sum_{{\cal C}}^\star e^{ \sqrt{\frac{\beta^2 M (q_2 + c_2 + c_{12})}{2}} \hat y_{\cal C}} \Bigg) \Bigg]} \right]^2 \right \} \\
& + n_1 n_2 \left \{ \overline{\Bigg[\ln \Bigg( \sum_{{\cal C}}^\star e^{ \sqrt{\frac{\beta^2 M (q_1 + c_1 + c_{12})}{2}} \hat x_{\cal C}} \Bigg) \Bigg]
\Bigg[ \ln \Bigg( \sum_{{\cal C}}^\star e^{ \sqrt{\frac{\beta^2 M (q_2 + c_2 + c_{12})}{2}} \hat y_{\cal C}} \Bigg) \Bigg]}^{(\star)} \right . \\
& \qquad \qquad \qquad \qquad \qquad
- \left . \overline{\Bigg[\ln \Bigg( \sum_{{\cal C}}^\star e^{ \sqrt{\frac{\beta^2 M (q_1 + c_1 + c_{12})}{2}} \hat x_{\cal C}} \Bigg) \Bigg]} \, 
\overline{\Bigg[ \ln \Bigg( \sum_{{\cal C}}^\star e^{ \sqrt{\frac{\beta^2 M (q_2 + c_2 + c_{12})}{2}} \hat y_{\cal C}} \Bigg) \Bigg]} 
\right \} \, ,
\eal
\end{equation}
where the average 
$\overline{[ g(\vec{\hat x}_{\cal C},\vec{\hat y}_{\cal C})]}^{(\star)}$ is defined over the gaussian measure:
\[
\overline{[ g(\vec{\hat x}_{\cal C},\vec{\hat y}_{\cal C})]}^{(\star)} \equiv \int \prod_{\cal C}^\star \left[ {\rm d} \hat x_{\cal C} \, {\rm d} \hat y_{\cal C} 
\frac{1}{2 \pi \sqrt{1 - \gamma^2}}
\, e^{- \frac{1}{1 - \gamma^2} \left( \frac{\hat x_{\cal C}^2}{2} + \frac{\hat y_{\cal C}^2}{2} -  
\gamma \hat x_{\cal C} \hat y_{\cal C} \right)} \right]
g(\vec{\hat x}_{\cal C},\vec{\hat y}_{\cal C}) \, .
\]
At this point we should find the saddle-point expressions of $q_1$, $q_2$ and $q_{12}$ that extremize $A [q_{ab}^{[1]},q_{ab}^{[2]},q_{ab}^{[12]}] $, Eq.~(\ref{eq:A2}), insert these expressions back into Eqs.~(\ref{eq:A2}), (\ref{eq:Z1Z2}), and (\ref{eq:Z0}), and finally determine ${\cal S}_{\rm rep} (c_1,c_2,c_{12})$. It is however important to remember that in order to obtain the second cumulant of the effective Hamiltonian we do not need the whole expression of ${\cal S}_{\rm rep} (c_1,c_2,c_{12})$, but only the terms of order $n_1 n_2$. It is then convenient to expand the saddle-point solutions of the overlaps in powers of $n_1$ and $n_2$. To compute the second cumulant we will only need to expand the overlap as in Eq.~(\ref{eq:q_exp}) of the main text. It is also convenient to define the following functions:
\[
\beal
& L_1 (\vec{z}_{\cal C}) \equiv \ln \Bigg( \sum_{{\cal C}}^\star e^{ \sqrt{\frac{\beta^2 M (q_1 + c_1 + c_{12})}{2}} z_{\cal C}} \Bigg) \, , 
& L_2 (\vec{z}_{\cal C}) \equiv \ln \Bigg( \sum_{{\cal C}}^\star e^{ \sqrt{\frac{\beta^2 M (q_2 + c_2 + c_{12})}{2}} z_{\cal C}} \Bigg) \, , \\
& K_1 (\vec{z}_{\cal C}) \equiv \frac{\sum_{{\cal C}}^\star z_{\cal C} \, e^{ \sqrt{\frac{\beta^2 M (q_1 + c_1 + c_{12})}{2}} z_{\cal C}}}
{\sum_{{\cal C}}^\star e^{ \sqrt{\frac{\beta^2 M (q_1 + c_1 + c_{12})}{2}} z_{\cal C}}} \, , 
& K_2 (\vec{z}_{\cal C}) \equiv \frac{\sum_{{\cal C}}^\star z_{\cal C} \, e^{ \sqrt{\frac{\beta^2 M (q_2 + c_2 + c_{12})}{2}} z_{\cal C}}}
{\sum_{{\cal C}}^\star e^{ \sqrt{\frac{\beta^2 M (q_2 + c_2 + c_{12})}{2}} z_{\cal C}}} \, , \\
& \textrm{such that~~} \frac{{\rm d} L_{1,2} (\vec{z}_{\cal C})}{{\rm d} q_{1,2}} = \frac{1}{2} \sqrt{\frac{\beta^2 M}{2(q_{1,2} + c_{1,2} + c_{12})}} 
K_{1,2} (\vec{z}_{\cal C}) \, . &
\eal
\]
In terms of these functions, we get
\begin{equation} \label{eq:A_n1n2}
\beal
A[q_1,q_2,q_{12}] & \approx \frac{\beta^2 M}{4} \left [ - n_1 \frac{q_1^2}{2} - n_2 \frac{q_2^2}{2} + n_1 n_2 q_{12}^2
+ n_1 (1 - c_1 - c_{12} ) (q_1 + c_1 + c_{12}) + n_2 (1 - c_2 - c_{12} ) (q_2 + c_2 + c_{12}) \right] \\
& - n_1(1 - c_1 - c_{12}) \overline{L_1 (\vec{z}_{\cal C})} - n_2 (1 - c_2 - c_{12} ) \overline{L_2 (\vec{z}_{\cal C})} \\
& - n_1 n_2 (1 - c_1 - c_2 - c_{12}) \left( \overline{L_1 (\vec{x}_{\cal C}) L_2 (\vec{y}_{\cal C}) }^{(\star)}
- \overline{L_1 (\vec{z}_{\cal C})} \, \overline{L_2 (\vec{z}_{\cal C})} \right)
\eal
\end{equation}
The extremization of $A[q_1,q_2,q_{12}]$ with respect to $q_1$ gives
\begin{equation} \label{eq:q1}
\beal 
q_1 & = (1 - c_1 -c_{12}) \left[ 1 - \sqrt{\frac{2}{\beta^2 M (q_1 + c_1 + c_{12})}} \, \overline{K_1 (\vec{z}_{\cal C})} \right] \\
& - n_2 (1 - c_1 - c_2 - c_{12}) \left \{ \frac{2 \gamma^2}{\beta^2 M (q_1 + c_1 + c_{12}) (1 - \gamma^2)^2}
\overline{\left[ \sum_{{\cal C}}^\star \left( x_{\cal C}^2 + y_{\cal C}^2 - \frac{1 + \gamma^2}{\gamma} x_{\cal C} y_{\cal C} \right ) - N (1 - \gamma^2) 
\right] L_1 (\vec{x}_{\cal C}) L_2 (\vec{y}_{\cal C})}^{(\star)} \right . \\
& \qquad \qquad \qquad \left . + \sqrt{\frac{2}{\beta^2 M (q_1 + c_1 + c_{12})}} \left( 
\overline{K_1 (\vec{x}_{\cal C}) L_2 (\vec{y}_{\cal C})}^{(\star)} - \overline{K_1 (\vec{z}_{\cal C})} \, \overline{L_2 (\vec{z}_{\cal C})}
\right) \right \} \, .
\eal
\end{equation}
We obtain the same equation for $q_2$ by changing all indices $1 \leftrightarrow 2$. Finally, the saddle-point equation for $q_{12}$ reads
\begin{equation} \label{eq:q12}
q_{12} = (1 - c_1 - c_2 - c_{12}) \frac{2 \gamma^2}{\beta^2 M (q_{12} + c_{12}) (1 - \gamma^2)^2} \overline{\left[ N (1 - \gamma^2) - 
\sum_{{\cal C}}^\star \left( x_{\cal C}^2 + y_{\cal C}^2 - \frac{1 + \gamma^2}{\gamma} x_{\cal C} y_{\cal C} \right ) \right] 
L_1 (\vec{x}_{\cal C}) L_2 (\vec{y}_{\cal C})}^{(\star)} \, .
\end{equation}
Inserting the expansion~(\ref{eq:q_exp}) into Eqs.~(\ref{eq:q1}) and (\ref{eq:q12}) allows us to obtain $q_1^{[0,0]}$, $q_1^{[0,1]}$, $q_2^{[0,0]}$, $q_2^{[1,0]}$, and $q_{12}^{[0,0]}$ which, once inserted into Eq.~(\ref{eq:A_n1n2}), finally yield the second cumulant. As for the computation of the first cumulant, we show explicitly how this can be done for $M \gg 1$ and we keep only terms to second order in $c_1$, $c_2$, and $c_{12}$.

After expanding the functions $K_{1,2} (\vec{z}_{\cal C})$ in powers of $\epsilon_{1,2} = \sqrt{\frac{M \beta^2 (q_{1,2} + c_{1,2} + c_{12})}{2}}$ up to the sixth order, one obtains
\[
q_{1,2}^{[0,0]} \approx \frac{1}{2^M} + \frac{M \beta^2 -2}{2^{M+1}} (c_{1,2} + c_{12}) + \frac{M \beta^2 (M \beta^2 - 4)}{2^{M+3}} (c_{1,2} + c_{12})^2 + \ldots \, ,
\]
which, of course, coincides with the first two terms of Eq.~(\ref{eq:q0-Kac}) with $c \to c_{1,2} + c_{12}$. Inserting these solutions into Eq.~(\ref{eq:q12}) leads to  the expression of the saddle-point value of $q_{12}$ in powers of the concentrations (as above, we only consider the leading terms in $1/2^M$):
\[
q_{12}^{[0,0]} \approx \frac{1}{2^M} - \frac{1}{2^M}(c_1 + c_2) + \frac{M \beta^2 - 2}{2^{M+1}} c_{12} + \frac{M \beta^2}{2^{2M+1}} (c_1 + c_2)^2
+ \frac{M \beta^2(M \beta^2 - 4)}{2^{M+3}} c_{12}^2 - \frac{M \beta^2}{2^{M+1}} (c_1+c_2)c_{12} \, .
\]
From the above results we self-consistently find that $\gamma \ge 0$. The expressions of $q_{1,2}^{[0,0]}$ and $q_{12}^{[0,0]}$, when inserted into Eq.~(\ref{eq:q1}), yield the corrections of order $n_2$ (resp. $n_1$) to the saddle-point value of $q_1$ (resp. $q_2$), which actually turns out to be very small  
for large $M$. Up to the leading terms in $2^M$, we find
\[
q_{1}^{[0,1]} \approx \frac{M^2 \beta^4}{2^{4M+2}} - \frac{M^2 \beta^4}{2^{4M+2}} (c_1 + c_2) + 
\frac{M^2 \beta^4}{2^{3 M +1}} c_{12} - \frac{M^2 \beta^4}{2^{3 M +1}} (c_1 + c_2) c_{12}
+ \frac{M^2 \beta^4}{2^{2M+2}} c_{12}^2 + \frac{M^2 \beta^4 ( M \beta^2 - 1)}{2^{5 M + 3}} (c_1 + c_2) c_1 \, .
\]
An analogous expression for $q_{2}^{[1,0]}$ is obtained by changing $1 \leftrightarrow 2$. Finally, collecting all these results together into Eqs.~(\ref{eq:A2}) and using Eq.~(\ref{eq:FRS}) allows us to obtain the expression of the second cumulant of the effective Hamiltonian (up to the second order in the concentrations
$c_1$, $c_2$, $c_{12}$):
\[
\beal
\frac{{\cal S}_2[c_1,c_2,c_{12}]}{N} &= -\frac{1}{N} \lim_{n_1,n_2 \to 0} 
\frac{{\cal S}_{\rm rep}[c_1,c_2,c_{12}]}{n_1 
n_2} = \frac{\beta^2 M}{4} c_{12}^2 - \lim_{n_1,n_2 \to 0} \frac{A[q_1,q_2,q_{12}]}{n_1 n_2} \\
& \approx \frac{M \beta^2}{2^{2 M + 2}} - \frac{M \beta^2}{2^{2 M+ 1}} (c_1 + c_2) + \frac{M \beta^2}{2^M + 1} c_{12} + 
\frac{M \beta^2}{2^{2 M + 2}} (c_1 + c_2)^2\\ 
& \qquad + \frac{M \beta^2}{4} \Big( 1 + \frac{M \beta^2 - 4}{2^{M+1}}
\Big) c_{12}^2 + \frac{M \beta^2}{2^{2 M + 1}} c_1 c_2 - \frac{M \beta^2}{2^{M + 1}} (c_1 + c_2) c_{12} \\
& \approx 
\frac{M \beta^2}{4} \Big( 1 + \frac{M \beta^2 - 4}{2^{M+1}}
\Big) c_{12}^2 + \frac{M \beta^2}{2^{M+1}} (1 - c_1 - c_2) c_{12} \, ,
\eal
\]
where in the last line we have only kept terms up to $O(1/2^M)$. By re-expressing the concentrations via Eq.~(\ref{eq:c1c2c12}), we finally obtain the second cumulant of the effective action given in Eq.~(\ref{eq:S2_2MKREM}) the main text.

\section{Exact solution of the fully connected random-field + random-bond effective Ising model}
\label{app:REMFC-RFIMI}

The fully connected random-field + random-bond Ising model defined by the Hamiltonian in Eq.~(\ref{eq:Heff_REMFC}) can be solved exactly. We drop the spin-independent random term and we first simplify the problem by neglecting the cross-correlations between the random fields and the random bonds (see below and the companion paper\cite{paper2} for a test of this approximation) as well as the off-diagonal part of the random field correlation. As a result, on has to consider Gaussian distributed random variables with
\begin{equation}
\begin{split}
\overline{\d h_i \d h_j}^{(0)} & = \Delta_h \delta_{ij} \, , \\
\overline{\delta \! J_{2,ij} \delta \! J_{2,kl} }^{(0)} & = \Delta_J (\delta_{ik} \delta_{jl} + \delta_{il} \delta_{jk})\, .
\end{split}
\end{equation}
By using the replica trick and performing Hubbard-Stratonovich transformations, the partition function of the model can be written as
\begin{displaymath}
	\overline{Z^n} = e^{\frac{nN}{2} \left( \frac{\Delta_J}{2} + \Delta_h \right)} \left(\frac{N \Delta_J}{2 \pi} \right)^{\! n(n-1)/2} 
\int_{-\infty}^{+\infty} \prod_a \textrm{d} m_a 
\int_{-i\infty}^{+i\infty} \prod_a \textrm{d} \mu_a
\int_{-\infty}^{+\infty} \prod_{a \neq b} \textrm{d} q_{a b} \,
e^{N F[\{m_a, \mu_a, q_{a b}\}]} \, ,
\end{displaymath}
with
\begin{displaymath}
	F[\{m_a, \mu_a, q_{a b}\}] = \sum_a \bigg[ (H - \mu) m_a + \frac{J_2}{2} m_a^2 +
\frac{J_3}{3!} m_a^3 + \frac{J_4}{4!} m_a^4 \bigg ] 
	- \frac{\Delta_J}{4} \sum_{a \neq b} q_{a b}^2 + \ln Z_1 \, .
\end{displaymath}
The partition function of the single-site problem reads
\begin{displaymath}
Z_1 = \sum_{\{ \sigma_a \}} \exp \bigg[ 
	\sum_a \mu_a \sigma_a + \frac{1}{2} \sum_{a \neq b} \big( \Delta_J  q_{a b} + \Delta_h \big ) 
\sigma_a \sigma_b \bigg] \, .
\end{displaymath}
The saddle-point equations then provide
\begin{displaymath}
\begin{split}
m_a &= \langle \sigma_a \rangle \, \\
q_{ab} & = \langle  \sigma_a \sigma_b \rangle \, , \\
	\mu_a &= H + J_2 m_a + \frac{J_3}{2} m_a^2 + \frac{J_4}{6} m_a^3 \, ,
\end{split}
\end{displaymath}
where the averages are computed by using the single-site Hamiltonian. Taking the replica-symmetric (RS) ansatz, $m_a = m$, $q_{ab} = q$, $\mu_a = \mu$, we find
\begin{displaymath}
	Z_1 = e^{-\frac{n}{2} ( \Delta_J q + \Delta_h )} \int \mathcal{D}z 
\left[ 2 \cosh \left(z \sqrt{\Delta_J q + \Delta_h} + \mu \right) \right]^n \, ,
\end{displaymath}
where $\mathcal{D}z = e^{-z^2/2} {\rm d} z/\sqrt{2 \pi}$.
To the leading order in $N$ in the $n \to 0$ limit, the free energy per spin then reads
\begin{displaymath}
	f(m,q,\mu) = (H - \mu) m + \frac{J_2}{2} m^2 
+ \frac{J_3}{3!} m^3 + \frac{J_4}{4!} m^4  
	+ \frac{\Delta_J}{4} q^2 - \frac{\Delta_J}{2} q 
+ \int \mathcal{D}z \ln 2 \cosh \left( z \sqrt{\Delta_J q + \Delta_h} + \mu \right) \, .
\end{displaymath}
After taking the derivatives with respect to $m$, $\mu$, and $q$, we obtain the following self-consistent equations:
\begin{equation} \label{eq:REMFC-solution}
\begin{split}
	m &= \int \mathcal{D}z \tanh \left( J_2 m + \frac{J_3}{2} m^2 + \frac{J_4}{6} m^3 + H + z \sqrt{\Delta_J q + \Delta_h} \right) \, , \\
	q & = 1 - \frac{1}{\sqrt{\Delta_J q + \Delta_h}} \int \mathcal{D}z \, z \tanh \left( J_2 m + \frac{J_3}{2} m^2
	+ \frac{J_4}{6} m^3 + H + z \sqrt{\Delta_J q + \Delta_h} \right) \, ,
\end{split}
\end{equation}
which can be easily solved numerically.

The result found by using the approximate effective theory [\ie solving Eqs.~(\ref{eq:REMFC-solution}) by using the effective coupling constants and variances of the random terms given in Eqs.~(\ref{eq:HJ2J3}) and~(\ref{eq:HJ2J3-var})] is plotted in Fig.~\ref{fig:overlap} for $M=3$ (red curve, squares).  It shows a good quantitative agreement with the exact solution. 


Finally, one could wonder whether the higher-order correlations of the distributions of the random bonds and random fields play an important role. In order to check this, we have repeated the calculation, taking now into account more terms of the disorder distributions, namely the correlation between random fields on different sites and the correlation between random fields and random bonds [see Eq.~(\ref{eq:HJ2J3-var})]:
\[
	\overline{\delta h_i \delta J_{2,kj}} = \kappa \frac{\delta_{ij} + \delta_{ik}}{\sqrt{N}} \, .
\]
In this case the saddle-point equations read
\begin{equation} \label{eq:REMFC-solution-disorder}
\begin{split}
	m &= \int \mathcal{D}z \tanh \left[ (J_2 + 2 \kappa) m + \frac{J_3}{2} m^2 + \frac{J_4}{6} m^3 + H - \kappa q+ z 
\sqrt{\Delta_J q + \Delta_h + 2 \kappa m} \right] \, , \\
q & =  1 - \frac{1}{\sqrt{\Delta_J q + \Delta_h + 2 \kappa m}} 
	\int \mathcal{D}z \, z \tanh \left[ (J_2 + 2 \kappa) m + \frac{J_3}{2} m^2
	+\frac{J_4}{6} m^3 + H - \kappa q + z \sqrt{\Delta_J q + \Delta_h + 2 \kappa m} \right] \, .
\end{split}
\end{equation}
We have solved these equations numerically for $\kappa = M \beta^2(1 + 2^{2-M})/32$ 
and found no significant difference with respect to the case in which these higher-order correlations of the disorder distribution are neglected. (We will come back
to this point in the companion paper\cite{paper2}, see, {\it e.g.}, Fig.~5.)


\begin{thebibliography}{99}
\bibitem{quenched_barrier} 
Actually, for values of the overlap close to the barrier one, a quenched computation is needed, but even quantitatively, on the basis of previous results, one does not expect significant changes.
\bibitem{footnote_S0} 
Note that in the magnetic representation in terms of (hard) Ising variables, the role of the random energy ${\cal S}_0$ is to ensure that the variance $\mathcal S_2$ of the effective Hamiltonian is equal to zero in the zero-overlap (liquid) state.
\end{thebibliography}
\end{document}